\documentclass[universe,article,accept,pdftex,moreauthors]{Definitions/mdpi}
\usepackage{bm}

\pdfoutput=1
\firstpage{1} 
\pubvolume{1}
\issuenum{1}
\articlenumber{0}
\pubyear{2025}
\copyrightyear{2025}
\externaleditor{Firstname Lastname} 
\datereceived{ } 
\daterevised{ } 
\dateaccepted{ } 
\datepublished{ } 
\hreflink{https://doi.org/} 

\newcommand{\newtext}[1]{{\textcolor{black}{#1}}}

\Title{Classification of Symmetric Four-Body Dziobek Central Configurations and Application to the Earth--Moon System}

\Author{{Z}al{\'a}n Czirj{\'a}k $^{1,2}$\orcidA{}, B{\'a}lint {\'E}rdi $^{2}$ and Emese Forg{\'a}cs-Dajka $^{2,3,4,}${*}\orcidB{}}

\AuthorNames{Zal{\'a}n Czirj{\'a}k, B{\'a}lint {\'E}rdi and Emese Forg{\'a}cs-Dajka}

\address{%
$^{1}$ \quad {Research and Development Institute for Wildlife and Mountain Resources, Progresului 35B, 530240 Miercurea-Ciuc, Romania; czirjakzalan@icdcrm.ro} \\
$^{2}$ \quad {Department} 
 of Astronomy, Institute of Physics and Astronomy, {ELTE}  E\"otv\"os Lor\'and University, P\'azm\'any P\'eter {s}\'et\'any 1/A{,} 
 \mbox{H-1117 Budapest,}  Hungary; {b.erdi@astro.elte.hu}  
\\
$^{3}$ \quad {HUN-REN-SZTE}
 Stellar Astrophysics Research Group, Szegedi ú{t} 
 Kt. 766, H-6500 Baja, Hungary \\
$^{4}$ \quad {HUN-REN}  Research Centre for Astronomy and Earth Sciences, Konkoly Observatory, {MTA} 
 Centre of Excellence{,} 
 H-1121 Budapest, Hungary\\}
\corres{Correspondence: e.forgacs-dajka@astro.elte.hu}

\abstract{
Central configurations are fundamental equilibrium solutions of the Newtonian $n$-body problem and play a key role in understanding the structure and dynamics of gravitational systems. However, the classification and enumeration of such configurations remain incomplete in the four-body case, particularly for symmetric configurations.
In this work, we develop a framework for determining and classifying symmetric four-body Dziobek configurations. The method allows the explicit determination of the number of admissible configurations directly from the mass parameters, without requiring prior knowledge of their geometric structure. Combined with previously established semi-analytical relations, this approach provides a systematic characterization of symmetric configurations in terms of mass ratios.
As a physically relevant application, we apply the framework to the Earth--Moon system and determine the possible symmetric four-body central configurations involving Earth- and Moon-mass bodies and an additional object of arbitrary mass. We identify both isolated configurations and continuous families of equilibrium solutions, extending the concept of libration points to the four-body problem.
The presented semi-analytical approach contributes to the understanding of equilibrium structures in multi-body gravitational systems and provides a foundation for further studies in celestial mechanics, planetary dynamics, and spacecraft motion in complex gravitational environments.
}


\begin{document}

\section{Introduction}

The Newtonian $n$-body problem is one of the central and longstanding problems of celestial mechanics. It describes the motion of $n$ point-like bodies interacting through mutual gravitational attraction. Given masses $m_i$ and barycentric position vectors $\bm{r}_i$, their motion is governed by the system of differential {equations} 
\begin{equation}
\label{eq:nbody}
m_i\bm{\ddot{r}}_i = k^2\sum_{j\neq i}\frac{m_im_j}{r_{ij}^3}(\bm{r}_j-\bm{r}_i) \quad 1\leq i\leq n,
\end{equation}
where $k$ is the Gaussian gravitational constant and $r_{ij}=|\bm{r}_i-\bm{r}_j|$ is the distance between bodies $i$ and $j$, \newtext{and the factor $k^2$ plays the role of the gravitational coupling constant in the chosen system of astronomical units. In particular, there is no additional hidden mass parameter in Equation~\eqref{eq:nbody}; the masses $m_i$ appear explicitly in the equation.} While the two-body problem admits a complete analytical solution, the general case for $n\geq 3$ does not. Consequently, identifying and characterizing special classes of solutions plays a fundamental role in understanding the structure and dynamics of gravitational systems.

Among such special solutions, central configurations occupy a distinguished position. In a central configuration, the mutual gravitational forces are proportional to the position vectors relative to the center of mass, resulting in accelerations of the form
\begin{equation}
\label{eq:cc}
\bm{\ddot{r}}_i = -\lambda\bm{r}_i \quad 1\leq i\leq n,
\end{equation}
where $\lambda>0$ is constant for all bodies. This condition leads to the defining equations of central configurations,
\begin{equation}
\label{eq:cc_crit}
\sum_{j\neq i}\frac{m_j}{r_{ij}^3}(\bm{r}_j-\bm{r}_i) = \Lambda\bm{r}_i \quad 1\leq i\leq n,
\end{equation}
with $\Lambda=-\lambda/k^2$. \newtext{Here, $\lambda$ is not prescribed independently; for a given central configuration it is determined by the masses and geometry of the system. In the present formulation, this dependence is represented through the combined parameter $\Lambda$.} 
Central configurations generate homographic solutions of the $n$-body problem and correspond to equilibrium solutions in appropriately rotating reference frames~\citep{whittaker1964treatise,poincare1893methodes,szebehely1967theory,smale1970topology}.

Beyond their mathematical importance, central configurations have profound physical relevance. They determine the location of libration points, which play a crucial role in celestial mechanics and astrodynamics. These equilibrium structures form the dynamical framework underlying the stability and architecture of planetary systems~\citep{szebehely1967theory,murray1999solar}. They are also fundamental to the design of space missions, including spacecraft operating near Sun--Earth and Earth--Moon libration points, such as SOHO, JWST, and other observatories~\citep{gomez2001dynamics,koon2008dynamical,Dunham2023DDA....5420403D}. Furthermore, central configurations provide insight into the long-term evolution, stability, and formation of planetary and satellite systems, as well as exoplanetary \mbox{architectures~\citep{Laskar1993PhyD...67..257L,Morbidelli2002mcma.book.....M,murray1999solar}}. In multi-body environments, equilibrium configurations generalize classical libration points and help identify dynamically stable regions suitable for spacecraft placement or natural accumulation of material.

The systematic study of central configurations dates back to the pioneering works of Euler and Lagrange, who discovered the collinear and triangular solutions of the three-body problem~\citep{euler1767motu,lagrange1772essai}. Their stability properties, first analyzed by Gascheau, Routh, and others~\citep{gascheau1843examen,routh1875on,danby1964stability,erdi1977on}, revealed the intricate relationship between mass ratios, geometry, and dynamical stability. These classical results established central configurations as fundamental building blocks of gravitational dynamics. Despite extensive efforts over more than two centuries, a complete classification of central configurations remains an open problem, especially in the four-body case~\citep{saari2011central}. While several important families have been identified~\citep{moulton1910straight,pizzetti1904casi,lehmann1891ueber,leandro2003finiteness}, the general relationship between mass parameters and configuration geometry remains only partially understood.

Symmetric configurations represent a particularly important and tractable class of solutions. Among these, the symmetric four-body Dziobek configurations (S4BDC), first studied by~\citet{dziobek1900uber}, exhibit a strong analytical relationship between mass ratios and geometric parameters. In previous studies~\citep{erdi2016central,czirjak2019study}, we demonstrated that in such configurations the mass ratios are analytically linked to the geometry, allowing a systematic investigation of their structure. These results provide a foundation for developing semi-analytical methods capable of determining and classifying symmetric four-body central configurations.

The motivation for studying symmetric four-body central configurations is twofold. From a theoretical perspective, they extend the classical Euler and Lagrange solutions to higher-order systems and contribute to the understanding of equilibrium structures in gravitational dynamics. From a practical perspective, they provide insight into equilibrium configurations relevant to planetary systems, multi-body environments, and spacecraft dynamics. In particular, identifying such configurations in systems involving Earth- and Moon-mass bodies may help characterize equilibrium structures in the Earth--Moon system and provide a theoretical basis for future mission concepts and dynamical investigations in multi-body gravitational fields~\citep{gomez2001dynamics,koon2008dynamical,murray1999solar}.

\newtext{
The present study builds upon our previous results and develops a semi-analytical framework for systematically determining and classifying symmetric four-body Dziobek configurations from the mass parameters. In Section~\ref{sec:overview}, we summarize and refine results established in~\citep{erdi2016central,czirjak2019study}, including the angular parametrizations and the relation between masses and geometry. The new contribution of the present paper is a direct-problem framework which, in the concave symmetric case, makes it possible to determine the number of admissible central configurations directly from the mass parameters, without prior knowledge of the corresponding geometry or of the placement of the masses on the symmetry axis. In Section~\ref{sec:cc_em}, we apply the refined framework to the Earth--Moon system and determine the possible symmetric four-body central configurations involving Earth- and Moon-mass bodies. Finally, Section~\ref{sec:conclusion} summarizes the main results. Appendices~\ref{app:coeff}--\ref{app:caseD} provide additional semi-analytical and numerical details supporting the analysis.
}

\section{The Mathematics of S4BDC Angle-Based Models}
\label{sec:overview}

According to~\citet{wintner1941analytical}, the {term} 
 \textls[-15]{\textit{central configurations} was introduced} \mbox{by~\citet{laplace1789quelques}} while he was examining the Lagrangian solutions of the three-body problem. According to his mode of discussion, giving a solution for the problem consists of determining all possible geometrical configurations and mass arrangements for a predetermined set of masses (this is the so-called direct problem). The inverse problem is a reverse procedure, where one determines the masses for configurations with given shapes.

Determining all the central configurations with a set number of bodies with predetermined masses seems hopeless because these types of systems are self-similar, i.e., invariant under rotation, translation, and dilation. A more reasonable task is to use the self-similarity property to define equivalence classes and count them. Though, counting these classes still proves to be an intricate task partly because, in some cases, they distinguish between configurations with the same shape but different labelling of the bodies. Thus, it is more convenient to count affine classes, which are the same as equivalence ones but without labelling of the bodies. In this paper, we define the number of central configurations as the number of affine classes.

In the following subsections, we first briefly overview our previous studies on the inverse problem (Section~\ref{sec:inv_prob}), and then utilize our results to solve the direct problem (Section~\ref{sec:prob}).

\subsection{Overview of the Inverse Problem}
\label{sec:inv_prob}

In this subsection, we sum up our results on the inverse problem described in~\citet{erdi2016central} and~\citet{czirjak2019study}.

In~\citep{czirjak2019study} we proved that in the S4BDC there is at least one axis of symmetry. Thus, to solve the inverse problem, one needs to study two cases:
\begin{enumerate}[label=(\alph*),ref=(\alph*)]
 \item \label{case:a} when there are no bodies on the axis of symmetry, 
 \item \label{case:b} when there are two bodies on the axis of symmetry.  
\end{enumerate}

A full description of {Case} 
 \ref{case:a} is given in~\citep{czirjak2019study}, while that of Case \ref{case:b} in~\citep{erdi2016central}.

\begin{figure}[H]

\includegraphics[width=\linewidth]{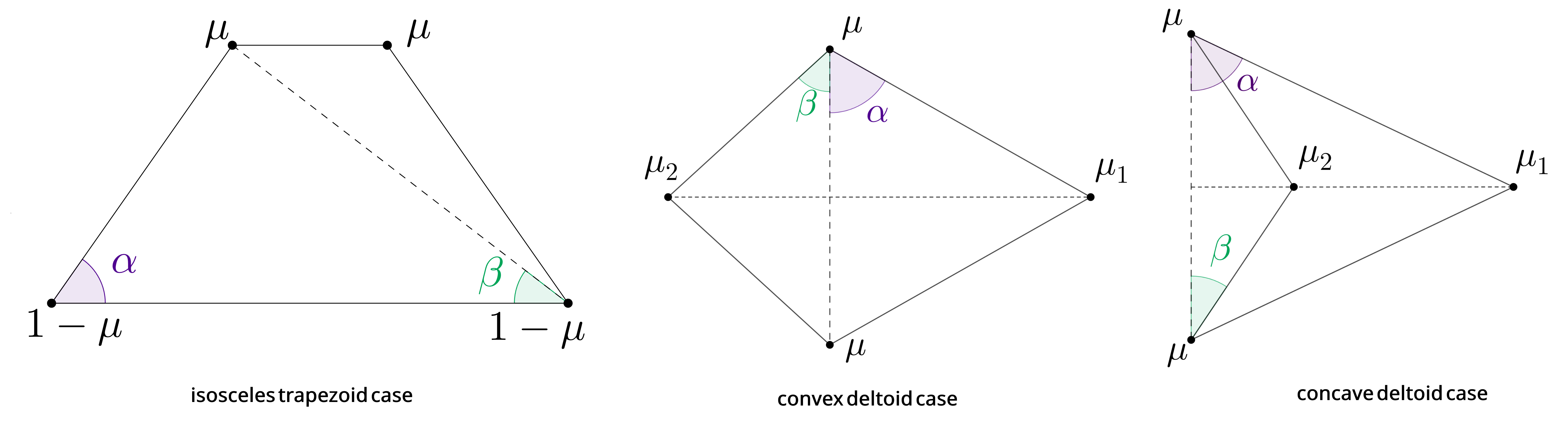}
\captionof{figure}{An isosceles trapezoidal configuration where neither body is located on the axis of symmetry. 
Two bodies are located on the axis of symmetry if the configuration is convex or concave and of deltoid type.
}
\label{fig:frames}
\end{figure}

In {Case} 
 \ref{case:a}, the configuration is an isosceles trapezoid, as shown on the left side of Figure~\ref{fig:frames}. The bodies that are reflections of each other (with respect to the axis of symmetry) have identical masses. The shape of the configuration uniquely determines the mass arrangement. To calculate the non-dimensional mass $\mu$, one needs to solve the following system of Equations~\citep{czirjak2019study}:
\vspace{-6pt}

\begin{adjustwidth}{-\extralength}{0cm}
\begin{equation}
\label{eq:trapez_crit}
\begin{split}
& [\sin^3\beta-\sin^3(\alpha+\beta)][\sin^3\alpha-\sin^3(\alpha-\beta)]  
	+[\sin^3\alpha-\sin^3(\alpha+\beta)][\sin^3\beta-\sin^3(\alpha-\beta)]=0,\\
& \mu=\frac{1}{1+
       \displaystyle{\frac{\sin^2(\alpha+\beta)}
                          {\sin^2(\alpha-\beta)} \cdot
                     \frac{\sin^3\alpha-\sin^3(\alpha-\beta)}
	                  {\sin^3\alpha-\sin^3(\alpha+\beta)}
                    }
               }.
\end{split}
\end{equation}
\end{adjustwidth}

Taking into account the geometrical restrictions ($0^\circ<\beta<\alpha\leq90^\circ$), the first equation establishes a one-to-one correspondence between the angles $\alpha$ and $\beta$. So just one angle is enough to define the shape of the configuration. The mass $\mu$ can be determined from the second equation of (\ref{eq:trapez_crit}), after substituting there the solutions $\alpha$ and $\beta$ of the first equation.

After solving the first of Equations~(\ref{eq:trapez_crit}), the coefficient $\Lambda$ appearing in Equation~(\ref{eq:cc_crit}) can be calculated as follows~\citep{czirjak2019study}:

\begin{equation}
    \Lambda = -\sin^3\alpha - \sin^3\beta.
\end{equation}

Configurations in Case \ref{case:b} take convex or concave deltoid shapes, as shown in the middle and right sides of Figure~\ref{fig:frames}. The bodies separated by the axis of symmetry have the same mass. In this case, the shape of the configurations also uniquely determines the mass arrangement. Using the notations from Figure~\ref{fig:frames}, the explicit formulas for the non-dimensional masses $\mu_1$, $\mu_2$ and $\mu$ are~\citep{erdi2016central}: 

\begin{equation}
\label{eq:deltoid_mu} 
\mu_1 = \frac{b_0(b_1+a_0-b_0)}{a_0b_1 + a_1b_0 -a_1b_1}, \quad \mu_2 = \frac{a_0(a_1+b_0-a_0)}{a_0b_1 + a_1b_0 -a_1b_1}, \quad \mu = \frac{1}{2}(1-\mu_1-\mu_2).
\end{equation}

\begin{figure}[H]
  
 \hspace{-20pt} \includegraphics[width=0.7\linewidth]{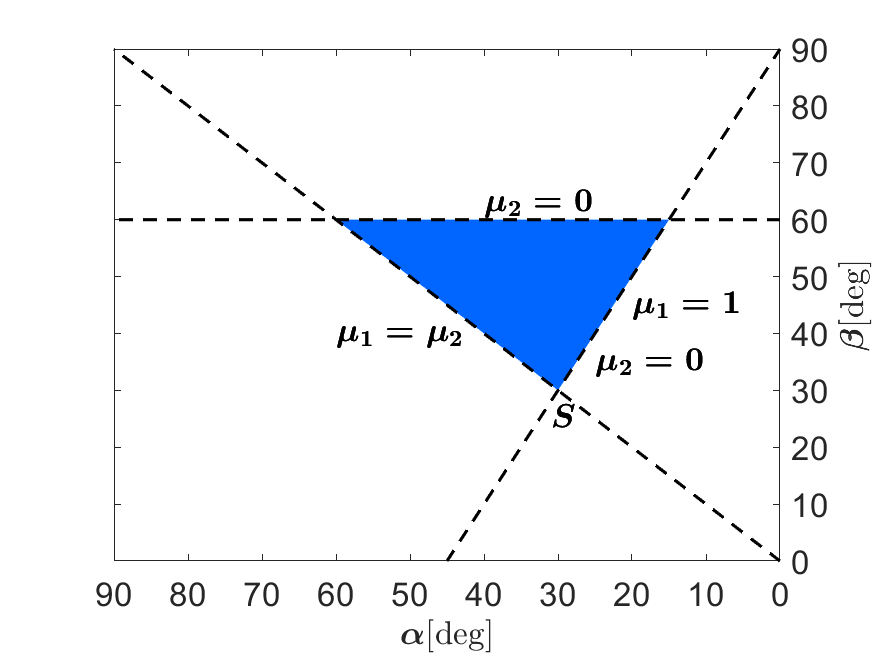}
  \caption{The blue area describes the domain of parameters that lead to acceptable convex solutions (if $\mu_2\leq\mu_1$). {The dashed lines indicate the critical points where the parameters result in the masses shown in the figure. The point} 
   $S$ is a singular point, where the mass functions given by Equation~(\ref{eq:deltoid_mu}) are undefined, instead $\mu_1+\mu_2=1$ holds with arbitrary masses.}
  \label{fig:convex_domain}
\end{figure}

\begin{figure}[H]
  
  \includegraphics[width=0.7\linewidth]{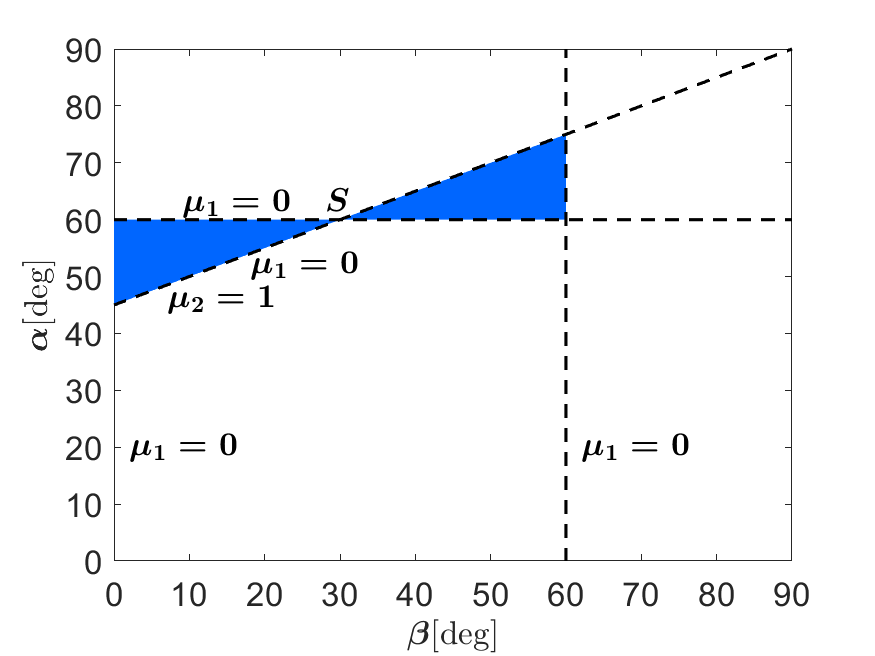}
  \caption{The blue triangle-shaped areas describe the domains of parameters that lead to acceptable concave solutions. On the left are the first, on the right are the second case solutions. {The dashed lines indicate the critical points where the parameters result in the masses shown in the figure. The point} 
 $S$ is a singular point, where the mass functions given by Equation~(\ref{eq:deltoid_mu}) are undefined, instead $3\mu_1+\mu_2=1$ holds with arbitrary masses.}
  \label{fig:concave_domain}
\end{figure}

The coefficients $a_0$, $a_1$, $b_0$ and $b_1$ are trigonometric polynomials of the angles $\alpha$ and $\beta$ (see Appendix \ref{app:coeff}). Beside the geometrical restriction on the angles ($0^\circ<\beta\leq\alpha<90^\circ$), there are restrictions on the masses as well ($0\leq\mu_1,\mu_2,\mu\leq1$ and $\mu_1+\mu_2+2\mu=1$). For acceptable solutions, these restrictions define so-called admissible regions on the parameter space (see Figures~\ref{fig:convex_domain} and \ref{fig:concave_domain}). Any pairs of the angles $\alpha$ and $\beta$ from the blue regions of Figures~\ref{fig:convex_domain} or \ref{fig:concave_domain} result in convex or concave configurations, respectively, with corresponding masses given by Equation~(\ref{eq:deltoid_mu}).

Note that the square configuration with four equal masses is the only common solution of Cases \ref{case:a} and \ref{case:b}.

\newtext{In Case \ref{case:b} once we have determined the non-dimensional masses $\mu_1$, $\mu_2$ and $\mu$ from the dimensionless angle parameters $\alpha$ and $\beta$ (the middle and right sides of Figure~\ref{fig:frames}) using Equation~(\ref{eq:deltoid_mu}). After calculating the dimensional form of the masses and the relative distances based on the angles mentioned above, the coefficient $\Lambda$ can be determined from Equation~(40) presented in~\citep{erdi2016central}. From this equation, an explicit formula for the coefficient $\Lambda$ can be derived, which depends on all the dimensional and dimensionless parameters mentioned above. For the sake of brevity in this summary, we will not go into details here.}

As a summary of this subsection, we emphasize that the shape of the configurations uniquely determines their mass arrangement. In the isosceles trapezoid case, solving Equation~(\ref{eq:trapez_crit}) will deliver the mass arrangement. Equation~(\ref{eq:deltoid_mu}) provide explicit formulae for calculating the masses of the deltoid-type central configurations.

\subsection{The Direct Problem}
\label{sec:prob}

In this subsection, we describe methods that are suitable to solve the direct problem of the S4BDC, utilizing the results on the inverse problem presented in Section~\ref{sec:inv_prob}.

Generally, studying the direct problem of central configurations of four bodies, a set of four positive values, representing the masses, is a priori given. In the problem of the S4BDC, there are some restrictions on the masses. In Case \ref{case:a}, there are two pairs of equal masses, while in Case \ref{case:b} there is at least one such pair (see Figure~\ref{fig:frames}).
 
Initially, there is no condition on the placement of the masses within the configurations. However, Equations (\ref{eq:trapez_crit}) and (\ref{eq:deltoid_mu}) determine the masses at a specific position within the configuration for it to be central (see Figure~\ref{fig:frames}). So using these formulae to calculate all possible solutions, this place-specific feature needs to be considered.

The set of masses for Case \ref{case:a} consists of two pairs of equal masses. Thus, the non-dimensional parameter $\mu$ describes the whole mass composition of the configuration (see the left panel of Figure~\ref{fig:frames}). In this case, the first of Equations~(\ref{eq:trapez_crit}) establishes a one-to-one correspondence between the angles $\alpha$ and $\beta$, similarly, the second of Equation~(\ref{eq:trapez_crit}) between $\mu$ and the angle pairs $(\alpha,\beta)$. So, Equation~(\ref{eq:trapez_crit}) has a unique solution. It can be determined knowing just one of the three parameters ($\mu$, $\alpha$ or $\beta$). So, in Case \ref{case:a}, the set of masses of the configuration uniquely determines its shape and vice versa.

Case \ref{case:b} involves convex and concave configurations (see the center and right panels of Figure~\ref{fig:frames}), each type having a different set of coefficients for the formulae (\ref{eq:deltoid_mu}). The set of masses contains maximum three different values because the bodies separated by the axis of symmetry are equal. So knowing two of the parameters $\mu_1$, $\mu_2$ and $\mu$ (see the center and right panels of Figure~\ref{fig:frames}) defines the mass composition of the configuration. 

In~\citep{czirjak2019study}, we have shown that there exists a unique convex configuration for each pair of the mass parameters $\mu_1$ and $\mu_2$, satisfying $\mu_1\geq\mu_2$. This condition excludes duplicated solutions due to the $180^\circ$ rotational symmetry around the center of mass of the system.

In~\citep{czirjak2019study}, we have also demonstrated that there exist between 0 and 2 concave type configurations for each pair of $\mu_1$ and $\mu_2$, depending on their values. In the concave case, there is no restriction for the masses with respect to their relative magnitude. Thus, their values can be interchanged. Furthermore, we studied the concave configurations considering $\mu_1$ and $\mu_2$ as place-specific mass parameters on the axis of symmetry, taking $\mu_1$ as the outer mass and $\mu_2$ as the inner one (see the right panels of Figure~\ref{fig:frames}). Here we consider the solutions of the concave case when the values of the mass parameters are interchangeable. Thus, let $m_1$ and $m_2$ be place-independent mass values, and we look for solutions when $\mu_1=m_1$, $\mu_2=m_2$, and $\mu_1=m_2$, $\mu_2=m_1$, where $\mu_1$ and $\mu_2$ are expressions given by Equation~(\ref{eq:deltoid_mu}). This problem is described in a general way by the following system of non-linear equations:

\vspace{-10pt}\begin{equation}
\label{eq:deltoid_prob}
\begin{split}
\mu_1(\alpha,\beta) &= m_i,\\
\mu_2(\alpha,\beta) &= m_j,
\end{split}
\end{equation}
where $i,j\in\{1,2\}$, $i\neq j$, $\mu_1$ and $\mu_2$ are expressions given by Equation~(\ref{eq:deltoid_mu}), and the angles $\alpha$ and $\beta$ are from within the blue regions in Figure~\ref{fig:concave_domain}. 

\begin{figure}[H]

  \includegraphics[width=.7\linewidth]{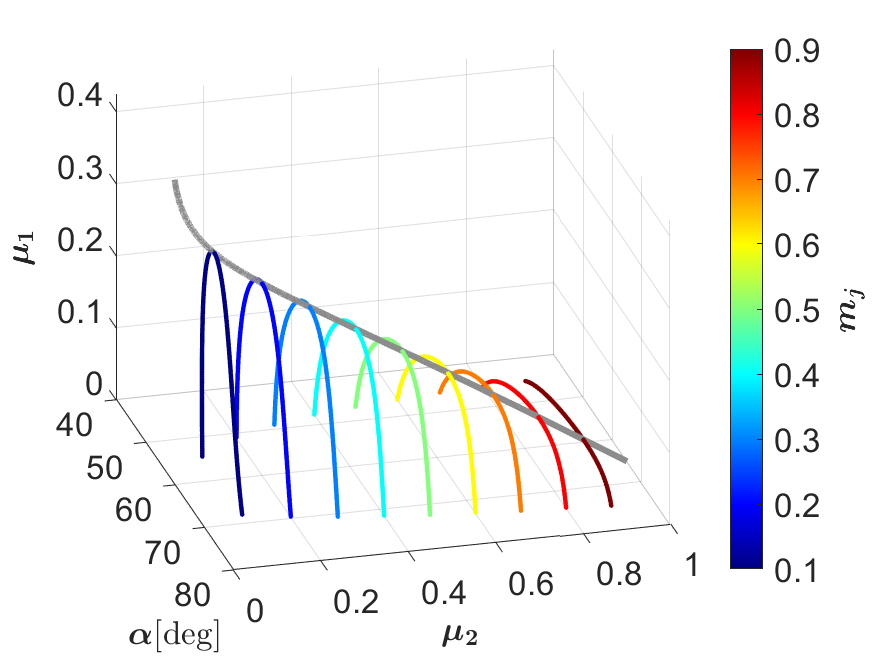}
  \caption{The $\mu_1(\mu_2=m_j)$ curves for different values of $m_j$. The value of $m_j$ is indicated by the colour of the curve and it is measured on the colour-scale. The gray curve represents the peak values $M_1(\mu_2)$.}
  \label{fig:concave_param}
\end{figure}

In~\citep{czirjak2019study}, we described a method for solving the system of Equation~(\ref{eq:deltoid_prob}). In the concave case, it begins with calculating the pairs of angle parameters $(\alpha, \beta)$ for which the second equation $\mu_2(\alpha,\beta)=m_j$ of Equation~(\ref{eq:deltoid_prob}) is satisfied. These pairs describe a continuous curve in the admissible angle domain (for examples of such curves see Figure 14 in~\citep{czirjak2019study}). Calculating $\mu_1$ along this curve  will lead to a characteristic curve $\mu_1(\mu_2(\alpha,\beta)=m_j)$. Figure~\ref{fig:concave_param} shows such characteristic curves for several values of $m_j$. Projecting this $\mu_1(\mu_2(\alpha,\beta)=m_j)$ curve onto the plane $(\alpha,\mu_1)$, its intersection(s) with the line $\mu_1=m_i$ will provide the solution of Equation~(\ref{eq:deltoid_prob}).

\begin{figure}[H]

  \includegraphics[width=.7\linewidth]{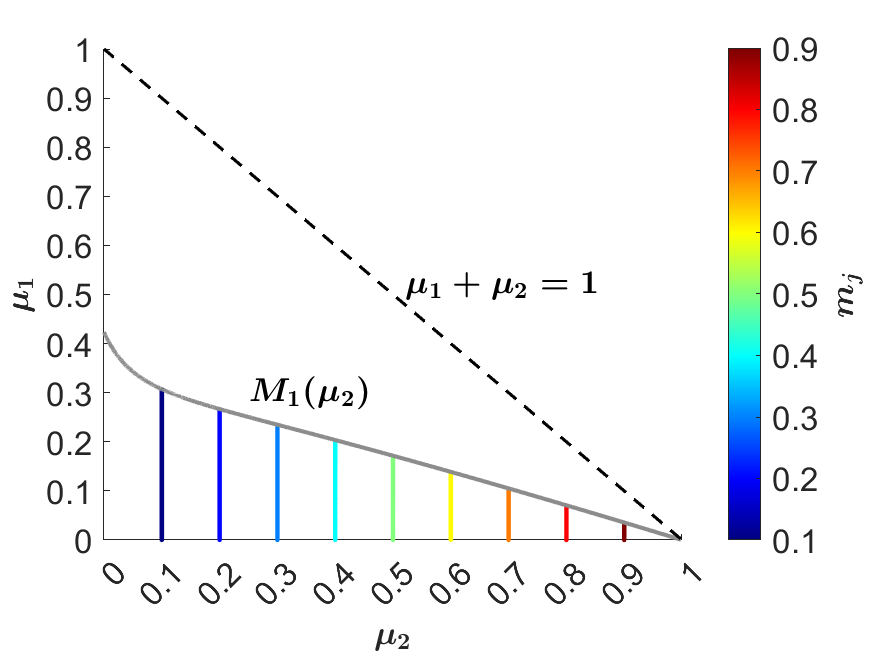}
  \caption{The curve $M_1(\mu_2)$ on the plane $(\mu_2,\mu_1)$, coloured by gray. {The dashed line indicates the boundary between possible and impossible configurations for the given mass parameters. Configurations below this line are possible, while those above it are impossible. The text on the line specifies the conditions for the configurations that are still possible.} 
  The coloured vertical lines correspond to the $\mu_1(\mu_2=m_j)$ curves in Figure~\ref{fig:concave_param}. The value of $m_j$ is measured on the colour-scale.}
  \label{fig:concave_mu}

\end{figure}

On the admissible angle domains (see Figure~\ref{fig:concave_domain}), the $\mu_1(\mu_2(\alpha,\beta)=c)$ function changes its monotonicity once (see Figure~\ref{fig:concave_param}) at a $M_1(\mu_2(\alpha,\beta)=c)$ maximum value, where $0\leq c\leq 1$ is a constant. The parameters $(\alpha,\beta,\mu_1,\mu_2)$ of the $M_1(\mu_2)$ peak values fulfill the following equation~\citep{czirjak2019study}:

\begin{equation}
\label{pr:concave_m1}
r\det(J_{qp}(\alpha,\beta))-p\det(J_{qr}(\alpha,\beta))-q\det(J_{rp}(\alpha,\beta)) = 0,
\end{equation}
where $\mu_1=q/r$, $\mu_2=p/r$ and
\begin{equation*}
J_{fg}(\alpha,\beta) =
\begin{pmatrix}
\frac{\partial f}{\partial\alpha} &\frac{\partial f}{\partial\beta}\\
\frac{\partial g}{\partial\alpha} &\frac{\partial g}{\partial\beta}
\end{pmatrix}
\end{equation*}
is the Jacobian matrix and $\det$ stands for the determinant. Figure~\ref{fig:concave_mu} shows the curve $M_1(\mu_2)$ on the plane $(\mu_2,\mu_1)$. Admissible values of $\mu_1$ and $\mu_2$ are below the line $\mu_1+\mu_2=1$; however, mass values only on and below the curve $M_1(\mu_2)$ lead to concave central configurations ($1$ or $2$ solutions, respectively). On the plane $(\mu_2,\mu_1)$ the curve $M_1(\mu_2)$ can be approximated by the following function:

\vspace{-6pt}\begin{equation}
\label{eq:approx_M1}
M_1(\mu_2) = 
\begin{cases} 
0.08258\cdot e^{-25.73\mu_2} + 0.3402\cdot e^{-1.227\mu_2}, & \mbox{if }  0\leq \mu_2 < 0.25034, \\ 
-0.03846\cdot\mu_2^2 -0.2859\cdot\mu_2 + 0.3241,  & \mbox{if }  0.25034 \leq \mu_2 \leq 1. 
\end{cases}
\end{equation}
{The} average precision of the functions (\ref{eq:approx_M1}) is $1.88\cdot10^{-5}$. The meaning of the critical value $\mu_2=0.25034$ in (\ref{eq:approx_M1}) is explained later.

On the $(\alpha,\mu_1)$ plane, the intersections of the curves $\mu_1(\mu_2(\alpha,\beta)=c)$ and the horizontal lines $\mu_1=c$, where $0\leq c\leq 1$, give the solutions of Equation~(\ref{eq:deltoid_prob}). The number of intersections can be assigned to each point of the $(\mu_2,\mu_1)$ plane. So examining the $(\mu_2,\mu_1)$ plane, for a given value of $\mu_2$, depending on the value of $\mu_1$, Equation~(\ref{eq:deltoid_prob}) has:
\begin{itemize}
 \item $0$ solution when $\mu_1>M_1(\mu_2)$,
 \item $1$ solution when $\mu_1=M_1(\mu_2)$,
 \item $2$ solutions when $\mu_1<M_1(\mu_2)$.
\end{itemize}

\begin{figure}[H]
  
  \includegraphics[width=.67\linewidth]{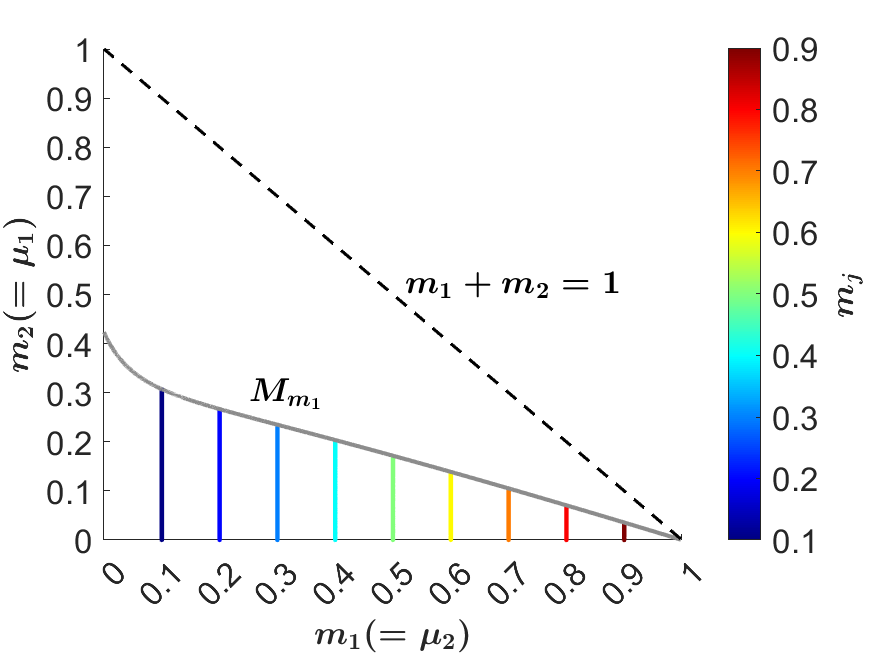}
  \caption{The $(m_1,m_2)$ parameter plane where the mass of the $\mu_1$-placed body (see the right panels of Figure~\ref{fig:frames}) is measured on the $m_2$ axis. The mass of the $\mu_2$-placed body is represented on the $m_1$ axis. The $\mu_1(\mu_2=m_1)$ curves from Figure~\ref{fig:concave_param} are the coloured vertical lines. The gray curve illustrates the $M_{m_1}=M_1(\mu_2=m_1)$ function. {The dashed line indicates the boundary between possible and impossible configurations for the given mass parameters. Configurations below this line are possible, while those above it are impossible. The text on the line specifies the conditions for the configurations that are still possible.} 
  }
  \label{fig:concave_m1}
\end{figure}

\begin{figure}[H]
  
  \includegraphics[width=.67\linewidth]{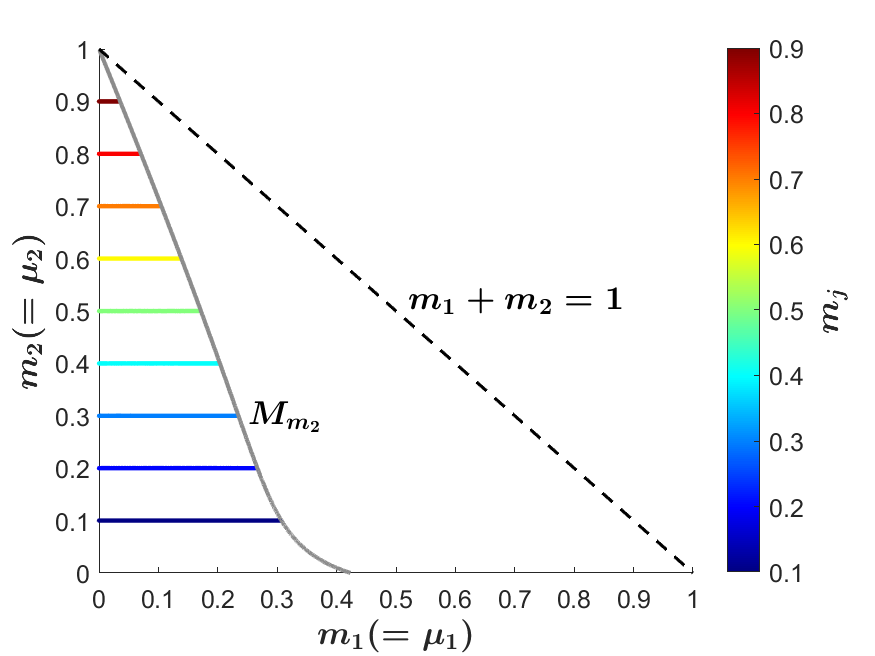}
  \caption{The $(m_1,m_2)$ parameter plane where the mass of the $\mu_1$-placed body (see the right panels of Figure~\ref{fig:frames}) is measured on the $m_1$ axis. The mass of the $\mu_2$-placed body is represented on the $m_2$ axis. The $\mu_1(\mu_2=m_2)$ curves from Figure~\ref{fig:concave_param} are the coloured horizontal lines. The gray curve illustrates the $M_{m_2}=M_1(\mu_2=m_2)$ function. {The dashed line indicates the boundary between possible and impossible configurations for the given mass parameters. Configurations below this line are possible, while those above it are impossible. The text on the line specifies the conditions for the configurations that are still possible.} 
  }
  \label{fig:concave_m2}
\end{figure}

For given parameters $m_1$ and $m_2$, all the possible concave type configurations are described by the combined solutions of both versions of Equation~(\ref{eq:deltoid_prob}), that is when $\mu_2=m_1$ and $\mu_2=m_2$. Let $M_{m_1}$ and $M_{m_2}$ denote the functions $M_1(\mu_2=m_1)$ and $M_1(\mu_2=m_2)$, respectively. These functions are shown in Figures~\ref{fig:concave_m1} and \ref{fig:concave_m2} on the parameter plane $(m_1,m_2)$. Figure~\ref{fig:concave_m1}, where $m_1$ and $m_2$ are measured on the horizontal and vertical axes, respectively, is actually a repetition of Figure~\ref{fig:concave_mu} with different notations. There are two solutions of Equation~(\ref{eq:deltoid_prob}) for masses below the curve $M_{m_{1}}$ in Figure~\ref{fig:concave_m1}, one solution on the curve $M_{m_{1}}$, and no (or zero) solution between the curve $M_{m_{1}}$ and the line $m_1+m_2=1$. Figure~\ref{fig:concave_m2}, similarly to Figure~\ref{fig:concave_m1}, where $m_1$ and $m_2$ are measured on the horizontal and vertical axes, respectively, is a mirrored version of Figure~\ref{fig:concave_mu} with respect to the diagonal $m_1=m_2$ with different notations. Thus, the curve $M_{m_{2}}$ in Figure~\ref{fig:concave_m2} separates regions with no or two solutions (right or left from the curve $M_{m_{2}}$), while for mass values along the curve $M_{m_{2}}$ there is one solution. The number of solutions of Equation~(\ref{eq:deltoid_prob}) is the sum of the solutions deduced from Figures~\ref{fig:concave_m1} and \ref{fig:concave_m2}, except for the cases when $m_1=m_2$. Switching equal masses on the axis of symmetry does not lead to a new (affine class) solution, so summation, in this case, is omitted. 

\begin{figure}[H]

  \includegraphics[width=.67\linewidth]{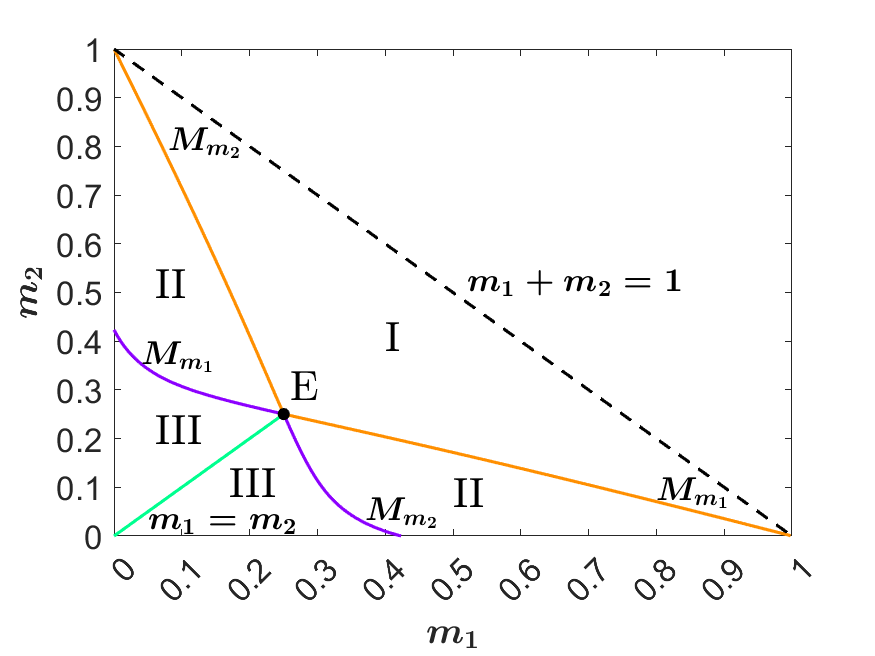}
  \caption{The three different types of subfields and their borders on the mass parameter plane.  A detailed description is provided in Table \ref{tab:concave_nr}. {The dashed line indicates the boundary between possible and impossible configurations for the given mass parameters. Configurations below this line are possible, while those above it are impossible. The text on the line specifies the conditions for the configurations that are still possible.} 
 }
  \label{fig:concave_diag}
\end{figure}

Figure~\ref{fig:concave_diag} is a combination of Figures~\ref{fig:concave_m1} and \ref{fig:concave_m2}. The curves $M_{m_1}$, $M_{m_2}$ (both having orange and indigo parts in the figure) and the (green) line $m_1=m_2$ define different borders of various types, dividing the admissible mass parameter plane into distinct areas. The curves $M_{m_1}$ and $M_{m_2}$ intersect each other in the point $E(m_1\sim0.25034,m_2\sim0.25034)$ on the line $m_1=m_2$. To each pair of coordinates $(m_1,m_2)$ in Figure~\ref{fig:concave_diag} there corresponds a specific number of solutions of Equation~(\ref{eq:deltoid_prob}). Table \ref{tab:concave_nr} summarizes these numbers. The counter function is described in Appendix \ref{app:counting}.

\vspace{-6pt}

\vspace{-6pt}

\begin{table}[H] 
\scriptsize
\caption{Number of concave deltoid-type central configurations as a function of the position of the point $(m_1,m_2)$ in the mass-parameter plane shown in Figure~\ref{fig:concave_diag}, where $m_1$ and $m_2$ are the masses located on the axis of symmetry of the configuration (see Figure~\ref{fig:concave_deltoid}). \label{tab:concave_nr}}
\begin{tabularx}{\textwidth}{cC}
\toprule
\textbf{Position of the Point \boldmath{$(m_1,m_2)$} in Figure~\ref{fig:concave_diag}} & \textbf{Number of Distinct Concave-Type Configurations} \\
\midrule
\newtext{within Region I} & 0\\
\midrule
\newtext{on the orange \textsuperscript{1} curves delimiting Region I, including point $E$} & 1\\
\midrule
\newtext{within Region II or on the green \textsuperscript{1} line segment separating the two parts of Region III} & 2\\
\midrule
\newtext{on the indigo \textsuperscript{1} curves separating Regions II and III} & 3\\
\midrule
\newtext{within Region III} & 4\\
\bottomrule
\end{tabularx}
\noindent{\footnotesize{\textsuperscript{1} \newtext{The colour labels are provided to facilitate identification in Figure~\ref{fig:concave_diag}.}}}
\end{table}

As a conclusion of this subsection, we emphasize that the set of masses describing the configuration generally does not uniquely determine its shape. The number of isosceles trapezoidal and convex deltoid central configuration is 1. So, their masses uniquely determine the shape of the configuration. However, the number of concave deltoid configurations is between 0 and 4. This number can be determined knowing the masses on the axis of symmetry (see Figure~\ref{fig:concave_deltoid}) using Figure~\ref{fig:concave_diag} with the help of Table \ref{tab:concave_nr}, or alternatively the function in Appendix \ref{app:counting}.

\subsection{\newtext{Literature Context and Contribution}}

\newtext{To conclude this section, we briefly position the present results within the existing literature and clarify the specific contribution of this paper. Section~\ref{sec:overview} summarizes and complements the results presented in~\citet{erdi2016central} and~\citet{czirjak2019study}. In~\citep{erdi2016central} we solved the inverse problem for planar axisymmetric deltoid central configurations in the four-body problem (4BP), while in~\citep{czirjak2019study} we discussed the other symmetric planar four-body cases and determined the number of configurations in all cases. Thus, the angular parametrizations, the convex/concave classification, and the mass--geometry relations used here are partly based on our earlier work.}

\newtext{The axisymmetric deltoid model presented in~\citep{erdi2016central} provides a common framework for several cases discussed by authors such as~\citet{dziobek1900uber},~\citet{long2002four}, \mbox{\citet{albouy2008symmetry}},~\citet{pina2010central}, and~\citet{deng2014planar}. The same framework has also proved useful in related problems:~\citet{kHovari2020axisymmetric} used it to characterize axisymmetric four-body central configurations with three equal masses;~\citet{gao2017equilibrium} and~\citet{alvarez2020overview} applied the corresponding solutions to the axisymmetric restricted five-body problem; and~\citet{veras2016relating} used it to study plausible planetary-system architectures of axisymmetric four-body central configurations containing two stars and one axis of symmetry.}

\newtext{The broader class of symmetric planar four-body central configurations has also been studied by~\citet{palmore1975classifying2},~\citet{simo1978relative},~\citet{meyer1988bifurcations},~\citet{leandro2003finiteness}, \mbox{\citet{shi2010classification}}, \citet{xie2012isosceles}, and~\citet{alvarez2013symmetric}. In this context, the results of~\citep{erdi2016central,czirjak2019study} are consistent with the classical classifications of~\citet{macmillan1932permanent}, \mbox{\citet{albouy1995recherches,albouy1996symmetric}}, and~\citet{perez2007convex}. Since the early 2000s, these deltoid-type configurations have often been referred to as ``kite configurations'' in the mathematical literature. In the present paper, however, we retain the term ``deltoid'' for consistency with our exposition.} 

\newtext{The new contribution of the present paper is not the rederivation of the inverse-problem results themselves, but their reformulation into a semi-analytical framework for the direct problem. In~\citep{czirjak2019study} the symmetric central configurations were grouped into three classes according to their shapes, and the number of configurations was determined for each class. In two of these classes the number of central configurations is one, whereas in the third class it varies between zero and two depending on the values and locations of the masses. In that case, the counting required the numerical solution of a relatively complicated extremum problem involving place-specific masses. In the present paper, building on those earlier results, we develop a semi-analytical counting procedure which determines the number of central configurations in this class directly from the corresponding mass values, without requiring prior knowledge of the placement of the masses on the symmetry axis or the explicit numerical solution of the associated extremum problem. This reformulation leads to a semi-analytical counting framework for the symmetric four-body Dziobek case considered here, in which the number of admissible configurations is obtained directly from the mass parameters rather than from a separate extremum analysis with place-specific masses.}

\newtext{For completeness, we also note that in the inverse-problem formulation the determination of the masses becomes independent of the coefficient $\Lambda=-\lambda/k^2$ introduced in Equations~(\ref{eq:nbody})--(\ref{eq:cc_crit}). In general, $\Lambda$ reflects both the mass distribution and the geometric scale of the configuration. However, it does not enter explicitly into the counting framework developed here. In Case~\ref{case:b}, $\Lambda$ may be determined based on Kepler’s third law~\citep{erdi2016central}, according to which~\citet{gauss1877theoria} determined the constant $k^2$. In Case~\ref{case:a}, by contrast, $\Lambda$ depends only on the shape of the system~\citep{czirjak2019study}.}

\section{Possible S4BDCs in the Earth--Moon System}
\label{sec:cc_em}

\newtext{In this section, we present a physically motivated application of the framework developed in Section~\ref{sec:overview} to the Earth--Moon system. Our aim is not to provide a complete dynamical or mission-design analysis, but to show how the semi-analytical classification introduced in the previous section can be applied to a concrete and astrophysically relevant mass hierarchy. In this sense, the Earth--Moon problem serves as a case study illustrating the possible symmetric four-body Dziobek configurations and their dependence on the mass of the additional body or bodies.}

\newtext{Central configurations are closely related to equilibrium structures in rotating multi-body systems, and in this sense they generalize the concept of libration points beyond the classical three-body setting. The Earth--Moon system, therefore, provides a natural example in which the framework of Section~\ref{sec:overview} can be illustrated in physically interpretable terms, even though questions of stability, detailed dynamics, and orbit design lie beyond the scope of the present paper.}

Let $m_1'$, $m_2'$, $m_3'$ and $m_4'$ denote the mass values of the four point-like bodies. Then let $m_1'$ and $m_2'$ be the masses of the Earth and Moon, respectively, that $m_1'=5.972 \cdot 10^{24}$ kg and $m_2'=7.349 \cdot 10^{22}$ kg. We use the average lunar distance $384,400$ km as the unit of length. Depending on the values of $m_3'$ and $m_4'$, we study the following {cases:} 
\begin{enumerate}[label=\Alph*.,ref=\Alph*.]
\item \label{em:a} $m_3'=m_1'$ and $m_4'=m_2'$ (or $m_3'=m_2'$ and $m_4'=m_1'$),
\item \label{em:b} $m_3'=m_4'>0$,
\item \label{em:c} $m_3'=m_2'$ and $m_4'>0$,
\item \label{em:d} $m_3'=m_1'$ and $m_4'>0$.
\end{enumerate}

In the following subsections, we characterize the main features of each case.

\subsection{Case \ref{em:a}}
\label{sec:case_a}

\begin{figure}[H]

\includegraphics[width=0.7\linewidth]{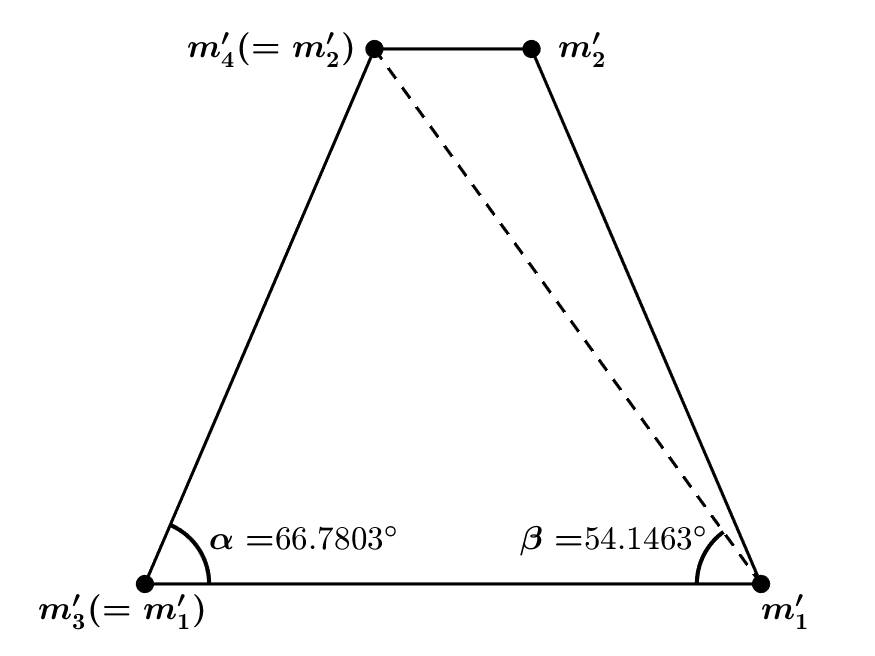}
\caption{Isosceles trapezoidal shape central configuration with bodies that have masses of the Earth and Moon.}
\label{fig:trapez1}
\end{figure}

In this case, the configuration takes an isosceles trapezoidal shape (see Figure~\ref{fig:trapez1}) and since $m_2'<m_1'$, thus, $\mu = m_2'/(m_1'+m_2')$. For the masses of the Earth and the Moon, $\mu=0.0122$. This type of configuration requires two pairs of bodies with equal masses. Thus, $m_3'=m_1'$ and $m_4'=m_2'$ (or $m_3'=m_2'$ and $m_4'=m_1'$) regardless of the labelling of $m_3'$ and $m_4'$. The only constraint is that the bodies with mass $m_1'$ and $m_2'$ in that order have to be on the long and short bases of the trapezoid.

Equation~(\ref{eq:trapez_crit}) has a unique solution for any of its three parameters (giving one, the other two can be uniquely determined). For $\mu=0.0122$, it follows that $\alpha=66.7803^\circ$ and $\beta=54.1463^\circ$ (see Figure~\ref{fig:trapez1}).

\subsection{Case \ref{em:b}}
\label{sec:case_b}

\begin{figure}[H]

\includegraphics[width=\linewidth]{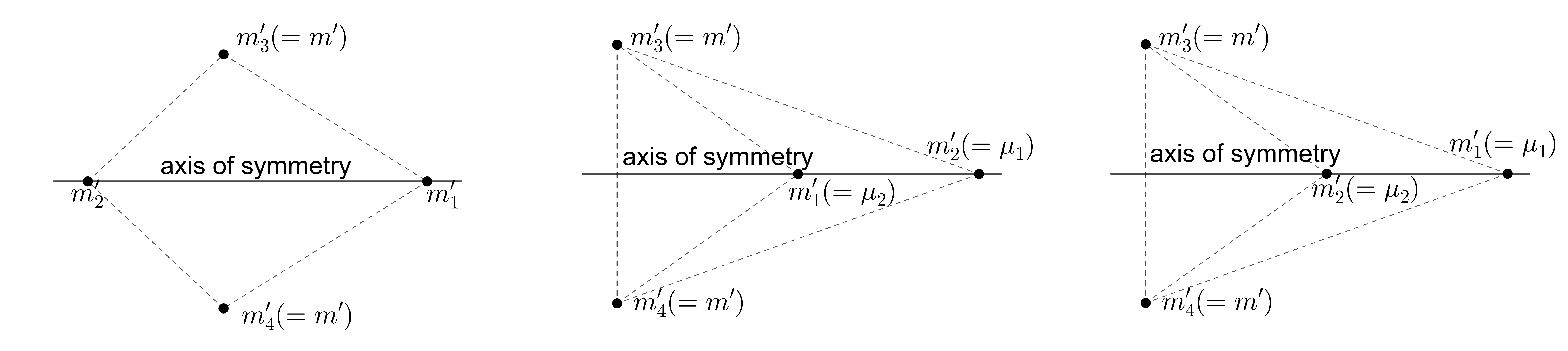}
\caption{Deltoid-type configurations in Case \ref{em:b}, where the bodies on the axis of symmetry have masses of the Earth and Moon.}
\label{fig:case_b}
\end{figure}

In this case, the deltoid-shaped configurations have the Earth's and Moon's mass bodies on the axis of symmetry (see Figure~\ref{fig:case_b}). We calculated the non-dimensional and place-independent mass parameters as
\begin{equation}
\label{eq:em_mass}
m_1=\frac{m_1'}{M},\quad m_2=\frac{m_2'}{M}, \quad m=\frac{m'}{M},
\end{equation}
where $M=m_1'+m_2'+2m'$ is the unit of mass, and $m'=m_3'=m_4'>0$, what we consider as an independent variable. We introduce the mass value $m''=m'/m_1'$ (the value of $m'$ in Earth's mass), to use it later as a reference in descriptions and illustrations.

\begin{figure}[H]
  
  \includegraphics[width=0.7\linewidth]{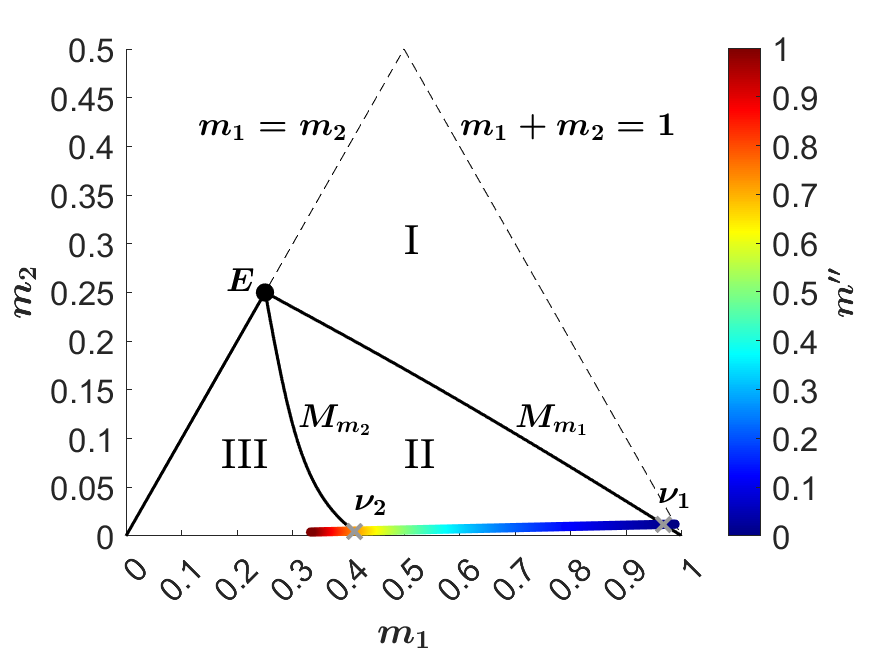}
  \caption{{This figure} 
   shows a portion of Figure~\ref{fig:concave_diag}, corresponding to the configurations examined in this case, bounded by dashed lines indicating the relevant relationships between the specified mass parameters. The coloured curve indicates the positions of the pairs $(m_1,m_2)$ for values of $m''$ between 0 and 1 Earth-mass according to the colour-scale. The number of the concave central configurations changes as the coloured curve goes through different regions, that is 0, 2, 4 in the regions I, II, III, respectively. At the points of intersection with the curves $M_{m_1}$ and $M_{m_2}$, there are 1 and 3 concave central configurations, respectively, whose mass parameters are denoted by the mark $x$ and named $\nu_1$ and $\nu_2$, respectively.}
  \label{fig:em_number}
\end{figure}

We studied configurations where the value of $m''$ changes between negligible and Earth's mass ($m''$ changes from zero to one). There is a unique convex configuration for any mass composition. We determined the number of the concave configurations with the help of Figure~\ref{fig:em_number}, where we displayed the points $(m_1,m_2)$, calculated by using Equation~(\ref{eq:em_mass}), and coloured them according to the value of $m''$. The coloured curve crosses the curve $M_{m_1}$ when $m''=\nu_1=0.0111$, the curve $M_{m_2}$ when $m''=\nu_2=0.7114$. Table \ref{tab:em_nr} lists the number of concave configurations (that changes between zero and four) depending on the value of $m''$.

\begin{table}[H]
\centering
\caption{The number of concave central configurations in Case \ref{em:b}, depending on the value of $m''$.}
\label{tab:em_nr}
\begin{tabularx}{\textwidth}{CC}
\toprule
\textbf{Condition} & \textbf{Number of Configurations} \\
\midrule
$ m''  < \nu_1$ & 0\\
$ m'' = \nu_1$ & 1\\
$ \nu_1 < m''  < \nu_2$ & 2\\
$ m'' = \nu_2$ & 3\\
$ \nu_2 < m''$ & 4\\
\bottomrule
\end{tabularx}
\end{table}

\begin{figure}[H]

\hspace{-55pt}\includegraphics[width=\textwidth]{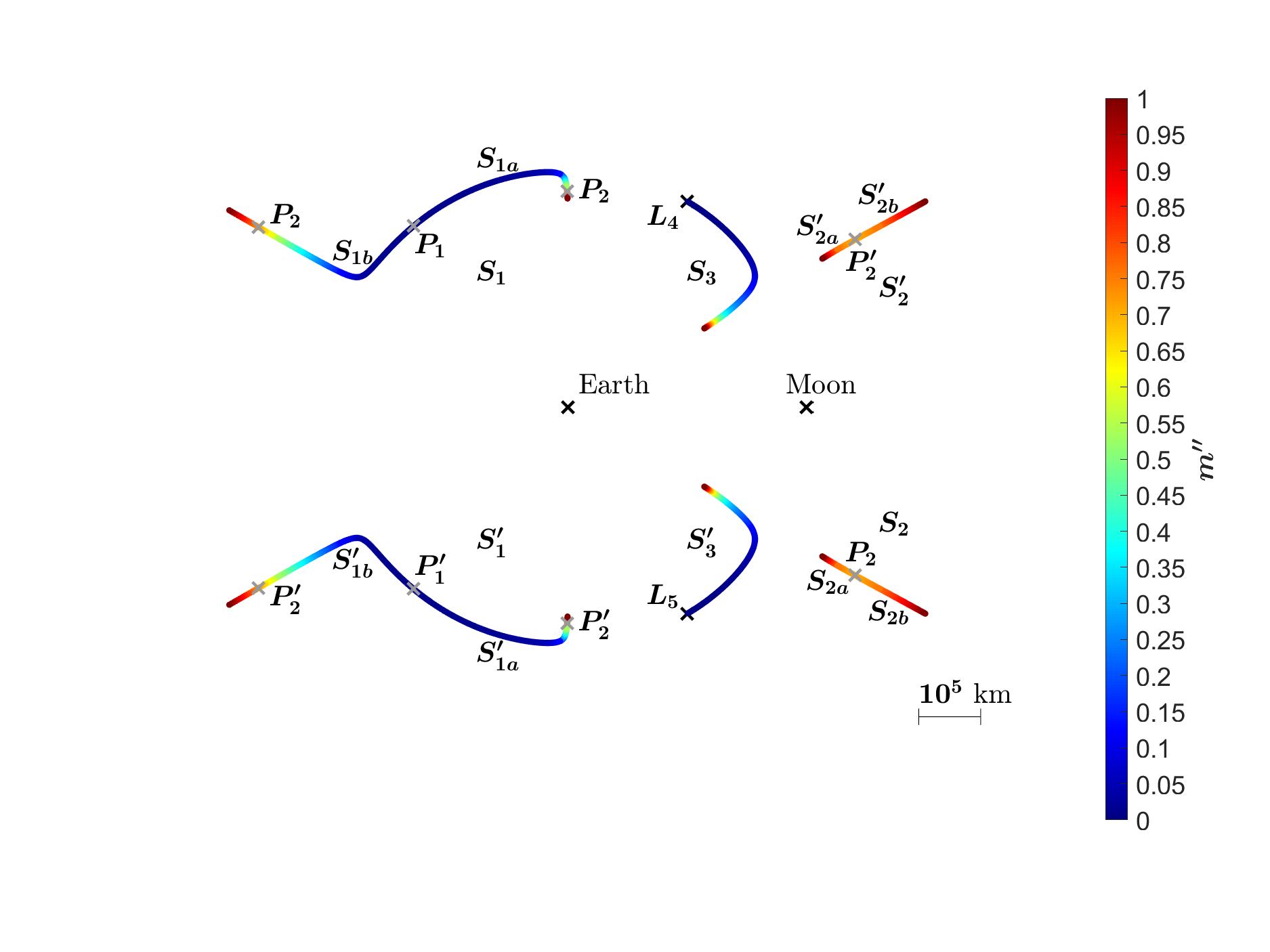}
\vspace{-12pt}\caption{The coloured curves $S_i$ and $S_i'$ represent the location of the equal-mass bodies in the Earth--Moon system in Case \ref{em:b}. These are the places, where they can form planar, symmetric central configurations with the Earth and the Moon. The colour-scale is linear in Earth-mass.}
\label{fig:em_map_big}
\end{figure}

We determined the locations of the equal-mass bodies $m_3'$ and $m_4'$ relative to the bodies $m_1'$ and $m_2'$ based on the value of $m''$, where the four bodies can form deltoid-type central configurations with the Earth--Moon axis being as the axis of symmetry. In Figure~\ref{fig:em_map_big}, the curves $S_i$ and $S_i'$ represent the locations of the equal-mass bodies {$m_3'$ and $m_4'$}, respectively. We calculated these curves by using Equations (\ref{eq:deltoid_mu}) and the formulae for the positions given in~\citep{erdi2016central}, and coloured them according to the value of $m''$. The points $P_1$ and $P_1'$ mark the locations of the equal-mass bodies when $m''=\nu_1$ ($P_1$ stands for $m_3'$, and $P_1'$ for $m_4'$), and $P_2$ (and $P_2'$) when $m''=\nu_2$ (similarly $P_2$ for $m_3'$ and $P_2'$ for $m_4'$). Note, that there are three $P_2$ (and $P_2$) points in accordance with Table \ref{tab:em_nr}. The curves $S_3$ and $S_3'$ represent the locations of the equal-mass bodies that lead to convex configurations. Equal-mass bodies on the curves $S_1$ and $S_1'$ or $S_2$ and $S_2'$ result in a concave configuration. The curves $S_1$ and $S_1'$ consist of two curves $S_{1a}$, $S_{1b}$ and $S_{1a}'$, $S_{1b}'$, respectively, that start from the points $P_1$ and $P_1'$, respectively, when $m''=\nu_1$ and end when $m''=1$ as the value of $m''$ grows continuously. The curves $S_1$ and the point $P_1$ are located on the side of the Lagrangian point $L_4$ with respect to the Earth--Moon axis. Similarly, the curves $S_2$ and $S_2'$ consist of two curves $S_{2a}$, $S_{2b}$ and $S_{2a}'$, $S_{2b}'$, respectively, that start from the points $P_2$ and $P_2'$, respectively, when $m''=\nu_2$ and end when $m''=1$ as the value of $m''$ grows continuously. The curves $S_2$ are located on the side of the Lagrangian point $L_5$ with respect to the Earth--Moon axis. The calculated configuration parameters of the endpoints of curves $S_i$ and $S_i'$ are summarized in Appendix \ref{app:caseB}.

\vspace{-12pt}

To conclude, there is a convex deltoid configuration for any mass arrangement, while concave configuration are only possible if $m''\geq \nu_1$. Table \ref{tab:em_nr} summarizes the number of concave-type configurations depending on the value of $m''$.

\subsection{Case \ref{em:c}}
\label{sec:case_c}

\begin{figure}[H]

\hspace{-16pt}\includegraphics[width=\linewidth]{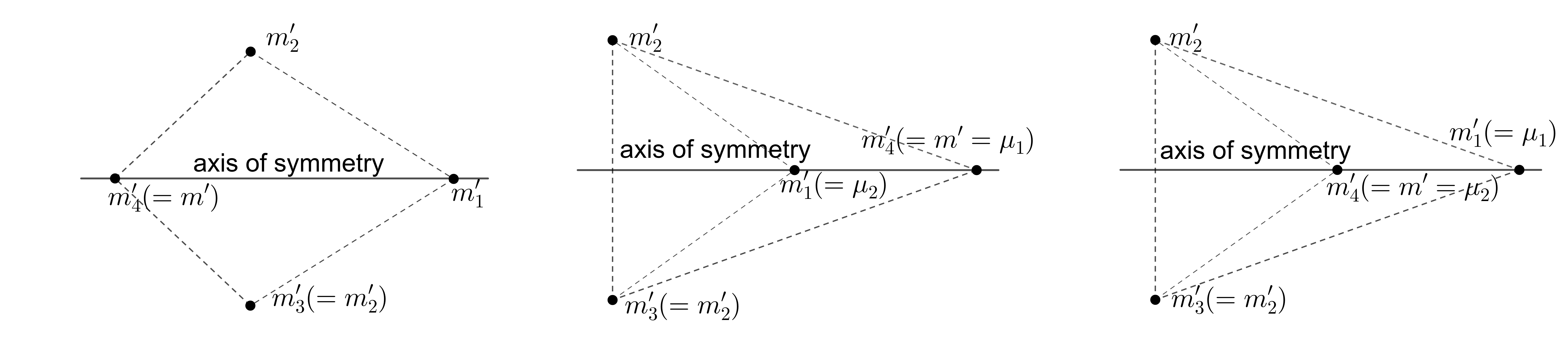}
\caption{Deltoid-type configurations in Case \ref{em:c}, where on the axis of symmetry, one body have the mass of the Earth, and the bodies that are separated by the axis of symmetry, have the mass of the Moon.}
\label{fig:case_c}
\end{figure}

In this case, the deltoid-shaped configurations have two Moon-mass bodies separated by the axis of symmetry, which houses two bodies, one with an Earth- and one with an arbitrary mass (see Figure~\ref{fig:case_c}). So, we calculated the non-dimensional and place-independent masses as 

\begin{equation}
\label{eq:e_mass}
m_1=\frac{m_1'}{M},\quad m_2=\frac{m'}{M}, \quad m=\frac{m_2'}{M},
\end{equation}
where $M=m_1'+2m_2'+m'$ is the unit of mass, $m_2'=m_3'$ and $m'=m_4'>0$ is an independent variable. Similarly to Case \ref{em:b}, we introduce the mass value $m''=m'/m_1'$.

\begin{figure}[H]
  
  \includegraphics[width=0.7\linewidth]{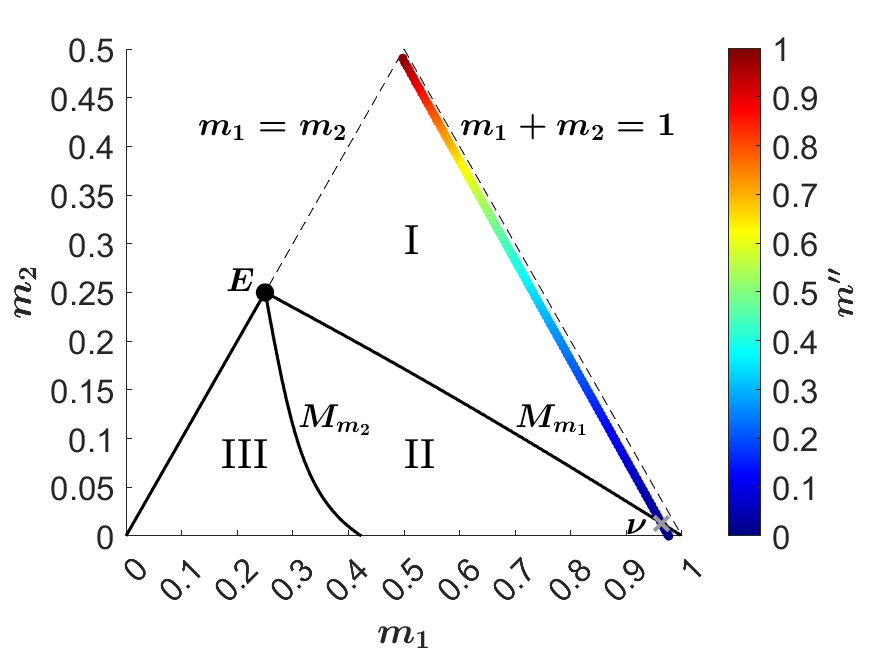}
  \caption{{This figure} 
   shows a portion of Figure~\ref{fig:concave_diag}, corresponding to the configurations examined in this case, bounded by dashed lines indicating the relevant relationships between the specified mass parameters. The coloured curve indicates the positions of the pairs $(m_1,m_2)$ for values of $m''$ between 0 and 1 Earth-mass according to the colour-scale. The figure corresponds to a part of Figure~\ref{fig:concave_diag}. The number of the concave central configurations changes as the coloured curve goes through different regions, that is 2, 0 in the regions II, I, respectively. At the point of intersection with the curve $M_{m_1}$, there is a concave central configuration for the mass parameter denoted by the mark $x$ and referred to as $\nu$.}
  \label{fig:e_number}
\end{figure}

Similarly to Case \ref{em:b}, we studied configurations for values of $m''$ between zero and one. There is a unique convex configuration for any mass composition. We deduced the number of concave configurations from Figure~\ref{fig:e_number}, based on the descriptions in Figure~\ref{fig:concave_diag} and Table \ref{tab:concave_nr}, where the points $(m_1,m_2)$, calculated by using Equation~(\ref{eq:e_mass}), are displayed and coloured according to the value of $m''$. The coloured curve crosses the curve $M_{m_1}$ when $m''=\nu=0.0137$. Table \ref{tab:e_nr} lists the number of concave configurations (that changes between zero and two) depending on the value of $m''$.

\begin{table}[H] 
  \centering
  \caption{The number of concave central configurations in Case \ref{em:c}, depending on the value of $m''$.}
  \label{tab:e_nr}
   \begin{tabularx}{\textwidth}{CC}
    \toprule
    \textbf{Condition} & \textbf{Number of Configurations} \\
    \midrule
    $ m''  < \nu $ & 2\\
    $ m''  = \nu $ & 1\\
    $ m''  > \nu $ & 0\\
    \bottomrule
  \end{tabularx}
\end{table}

\begin{figure}[H]

  \hspace{-47pt}\includegraphics[width=.67\linewidth]{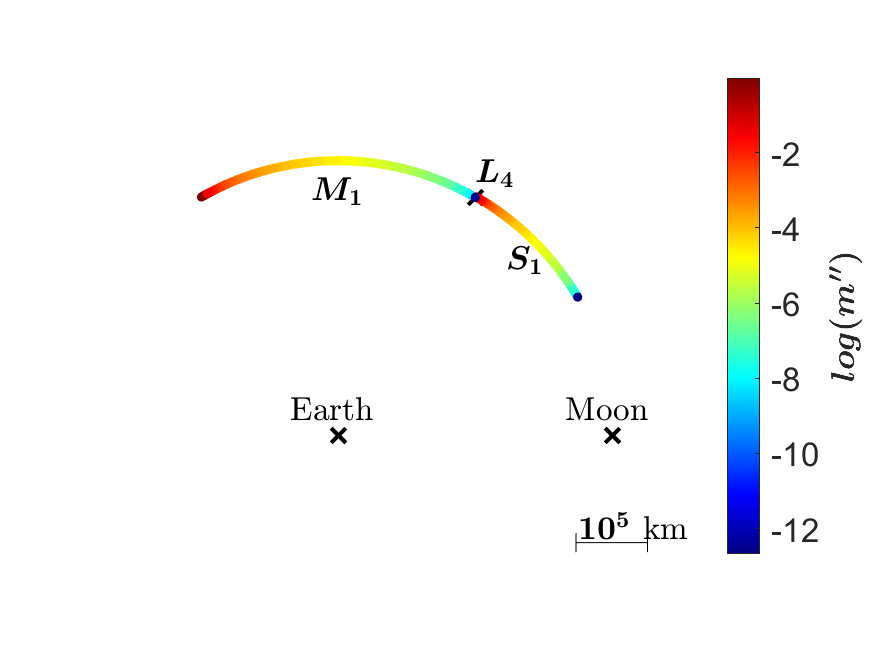}
  \captionof{figure}{{The} 
 coloured curves $M_1$ and $S_1$ represent the location of the bodies $m_3'$ and $m_4'$, respectively, to achieve convex configuration in Case \ref{em:c}. The curves $M_1$ and $S_1$ are coloured according to the value of $m''$. The colour-scale is logarithmic in Earth-mass.}
  \label{fig:e_convex}
\end{figure}

\begin{figure}[H]

 \hspace{-40pt} \includegraphics[width=.67\linewidth]{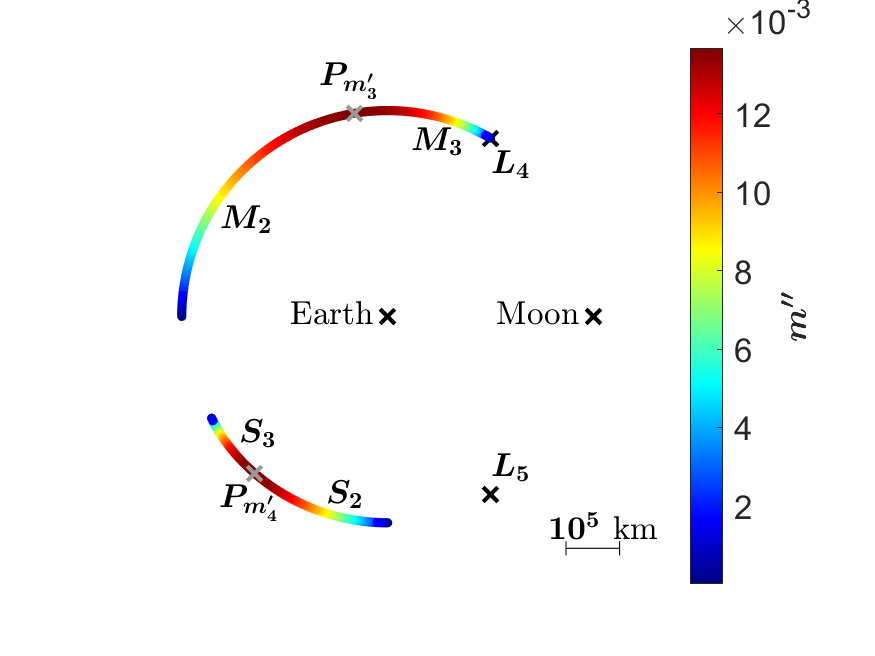}
  \captionof{figure}{{The}
 coloured curves $M_i$ and $S_i$ for $i=\{2,3\}$ represent the location of the bodies $m_3'$ and $m_4'$, respectively, to achieve concave configuration in Case \ref{em:c} when the Earth ($m_1'$) is the inner body. The curves $M_i$ and $S_i$ are coloured according to the value of $m''$. The colour-scale is linear in Earth-mass.}
  \label{fig:e_concave}
\end{figure}

We calculated the locations of the bodies $m_3'$ and $m_4'$ relative to the bodies $m_1'$ and $m_2'$ based on the value of $m''$, where the four bodies can form a deltoid-type central configuration containing the Earth--Moon axis (see Figure~\ref{fig:case_c}). In Figures~\ref{fig:e_convex} and \ref{fig:e_concave}, the curves $M_i$ and $S_i$, coloured according to the value of $m''$, indicate the positions of the bodies $m_3'$ and $m_4'$, respectively, with respect to $m_1'$ (Earth) and $m_2'$ (Moon).

In Figure~\ref{fig:e_convex}, the curves $M_1$ and $S_1$ represent the locations of the bodies $m_3'$ and $m_4'$, respectively, leading to convex configurations. As the value of the $m''$ grows, the body $m_4'$ heads toward the Lagrangian point $L_4$, while the body $m_3'$ departing from it is getting farther away. The colours do not match in Figures~\ref{fig:e_number} and \ref{fig:e_convex}. Both colour-scales takes values between zero and one, but it is linear in Figure \ref{fig:e_number}, and logarithmic in Figure~\ref{fig:e_convex} for better visibility.

In Figure~\ref{fig:e_concave}, the curves $M_i$ and $S_i$ for $i=\{2,3\}$ represent the locations of the bodies $m_3'$ and $m_4'$, respectively, leading to concave configurations. The points $P_{m_3'}$ and $P_{m_4'}$ represent the position of the bodies $m_3'$ and $m_4'$, respectively, when $m''=\nu$. In Figure~\ref{fig:e_concave}, the curves $M_i$ start with $m''=0$ and converge toward the point $P_{m_3'}$ as $m''$ grows toward $\nu$. Similarly, the curves $S_i$ begin at different points when $m''=0$ then head towards the point $P_{m_4'}$ as $m''$ grows to $\nu$. The colours do not match in Figures~\ref{fig:e_number} and \ref{fig:e_concave}. Both colour-scales are linear, but their range is 0--1 in Figure~\ref{fig:e_number}, and 0--0.001 in Figure~\ref{fig:e_concave}.

To conclude, there is one convex deltoid-type configuration for any value of $m''$, while concave configurations are only possible if $m''\leq \nu$. Table \ref{tab:e_nr} summarizes the number of concave-type configurations based on the value of $m''$. Table \ref{tab:e_param} (see Appendix \ref{app:caseC}) lists the configuration parameters calculated at the endpoints of the curves $M_i$ and $S_i$.

\subsection{Case \ref{em:d}}
\label{sec:case_d}

\begin{figure}[H]
\centering
\includegraphics[width=\linewidth]{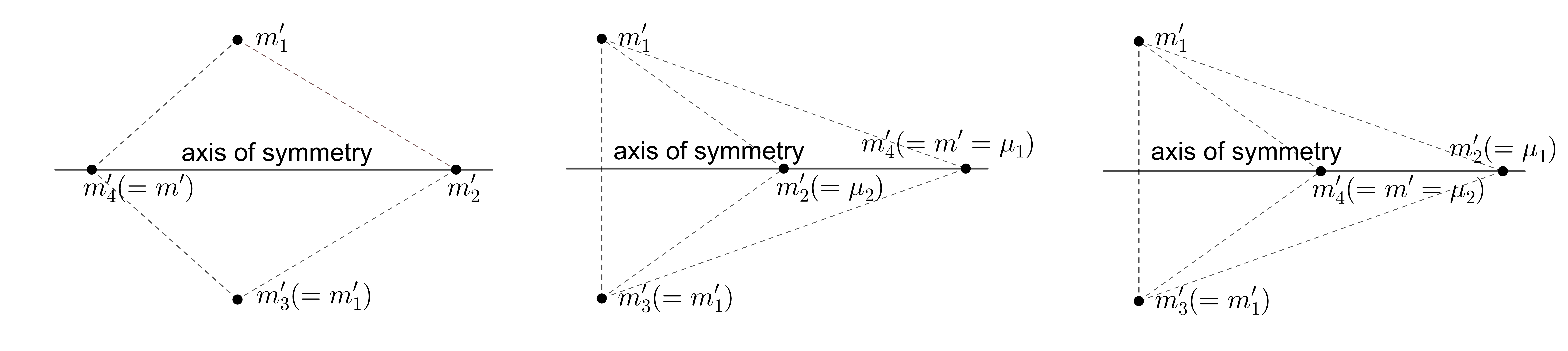}
\caption{Deltoid-type configurations in Case \ref{em:d}, where on the axis of symmetry, one body have the mass of the Moon, and the bodies that are separated by the axis of symmetry, have the mass of the Earth.}
\label{fig:case_d}
\end{figure}

In this case, the deltoid-shaped configurations have two Earth-mass bodies separated by the axis of symmetry, which contains two bodies, one with a Moon- and one with an arbitrary mass (see Figure~\ref{fig:case_d}). So, we calculated the non-dimensional and place-independent masses as

\begin{equation}
\label{eq:m_mass}
m_1=\frac{m'}{M},\quad m_2=\frac{m_2'}{M}, \quad m=\frac{m_1'}{M}.
\end{equation}
where $M=2m_1'+m_2'+m'$ is the unit of mass, $m_1'=m_3'$ and $m'=m_4'>0$ is an independent variable. Similarly to Cases \ref{em:b} and \ref{em:c}, we introduce the mass value $m''=m'/m_1'$.

\begin{figure}[H]
  
  \includegraphics[width=0.67\linewidth]{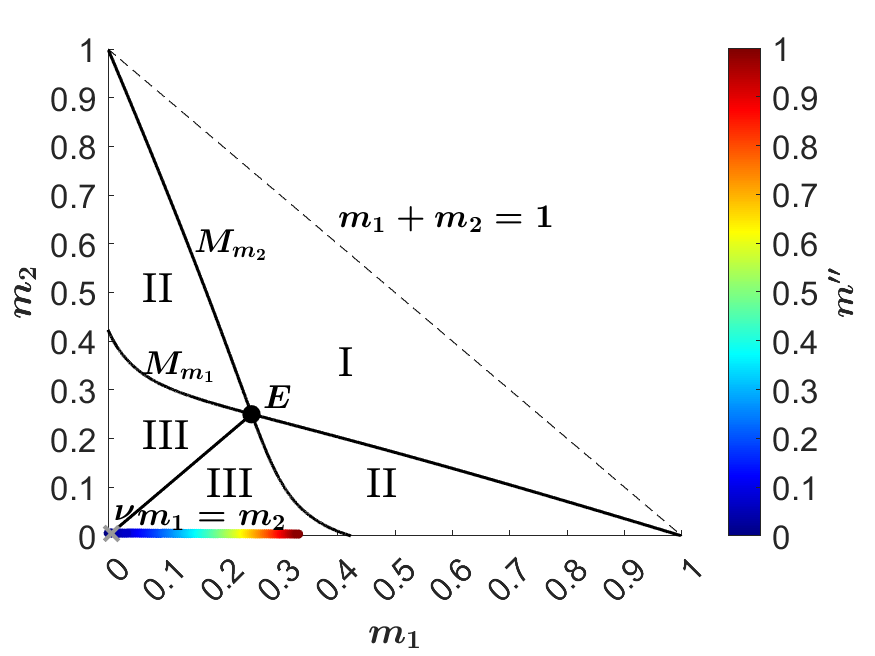}
  \captionof{figure}{{This figure} 
   illustrates the fully concave case shown in Figure~\ref{fig:concave_diag}. To avoid unnecessary repetition, the descriptions of the black curves and the dashed line can be found in the captions for Figure~\ref{fig:concave_diag} and the content of Table \ref{tab:concave_nr}. The coloured curve indicates the positions of the pairs $(m_1,m_2)$ for values of $m''$ between 0 and 1 Earth-mass according to the colour-scale. The number of the concave central configurations is four, since the coloured curve is inside the region III. Except, when it intersects the line $m_1=m_2$ at $\nu=0.0123$, which we denoted with the mark $x$ in the figure, then the number is two. }
  \label{fig:m_number}
\end{figure}

Similarly to Cases \ref{em:b} and \ref{em:c}, we studied configurations for values of $m''$ between zero and one. There is one convex configuration for all values of $m''$. To determine the number of concave configurations, we calculated the curve $(m_1,m_2)$, by using Equation~(\ref{eq:m_mass}), and coloured it according to the value of $m''$ (see Figure~\ref{fig:m_number}). Then we deduced the number of the concave configurations using the descriptions in Figure~\ref{fig:concave_diag} and Table \ref{tab:concave_nr}. In Figure~\ref{fig:m_number}, the coloured curve is inside the two regions of III, crossing the line $m_1=m_2$ at $m'' = \nu =0.0123$. So there are two different concave configurations when $m''=\nu$. For any other value of $m''$, there are four concave configurations.

\begin{figure}[H]
  
\hspace{-40pt}  \includegraphics[width=.7\linewidth]{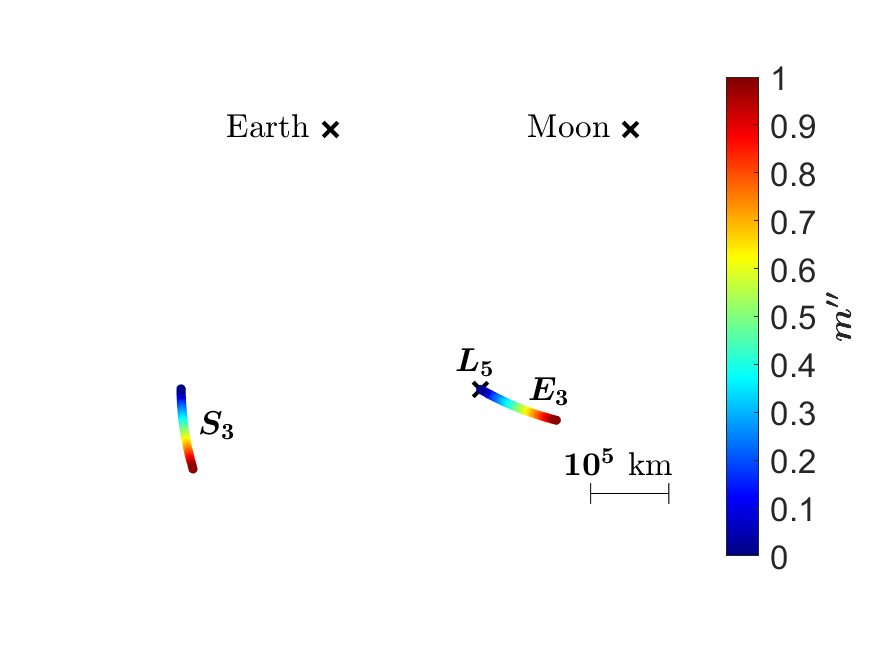}
  \captionof{figure}{The coloured curves $E_3$ and $S_3$ represent the location of the bodies $m_3'$ and $m_4'$, respectively, to achieve convex configuration in Case \ref{em:d}. The curves $E_3$ and $S_3$ are coloured according to the value of $m''$. The colour-scale is linear in Earth-mass.}
  \label{fig:m_convex}
\end{figure}

\begin{figure}[H]
  
\hspace{-46pt}  \includegraphics[width=.7\linewidth]{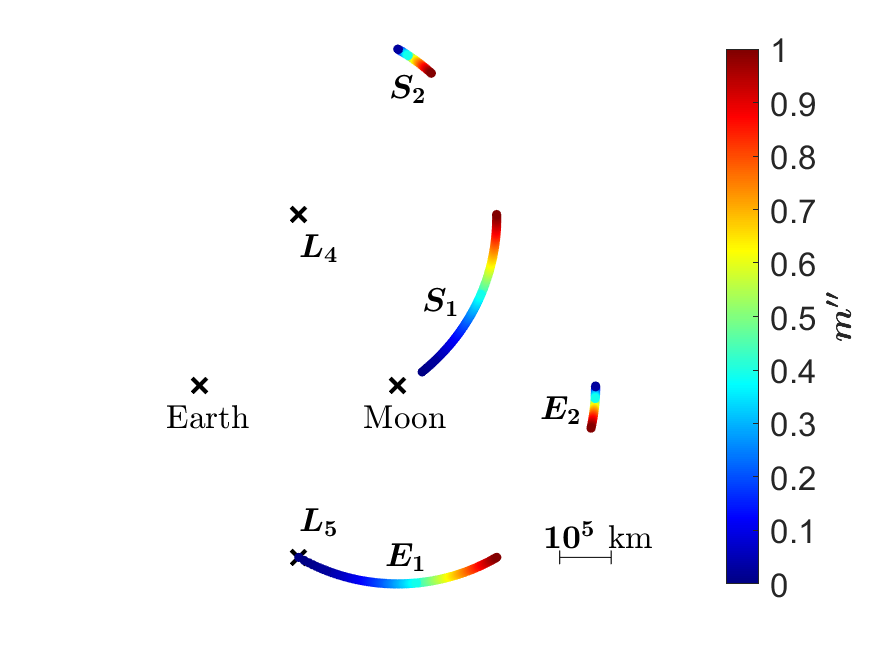}
  \captionof{figure}{The coloured curves $E_i$ and $S_i$ for $i=\{1,2\}$ represent the location of the bodies $m_3'$ and $m_4'$, respectively, to achieve concave configuration in Case \ref{em:d} when the inner body is the Moon ($m_2'$). The curves $E_i$ and $S_i$ are coloured according to the value of $m''$. The colour-scale is linear in Earth-mass.}
  \label{fig:m_m_concave}
\end{figure}

\begin{figure}[H]
  
 \hspace{-30pt} \includegraphics[width=0.7\linewidth]{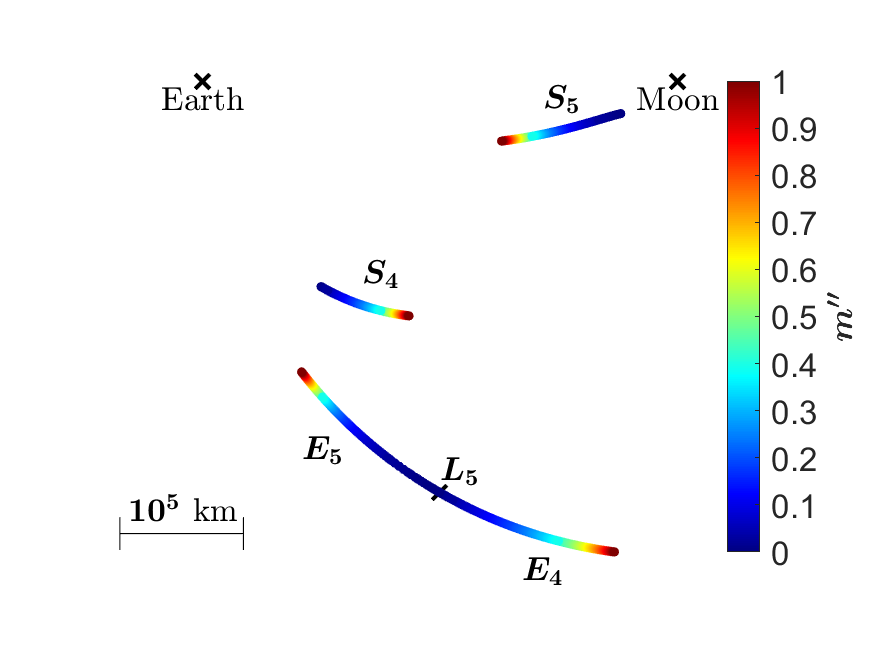}
  \captionof{figure}{The coloured curves $E_i$ and $S_i$ for $i=\{4,5\}$ represent the location of the bodies $m_3'$ and $m_4'$, respectively, to achieve concave configuration in Case \ref{em:d} when the inner body is the body $m_4'$. The curves $E_i$ and $S_i$ are coloured according to the value of $m''$. The colour-scale is linear in Earth-mass.}
  \label{fig:m_s_concave}
\end{figure}

Similarly to Cases \ref{em:b} and \ref{em:c}, we calculated the locations of the bodies $m_3'$ and $m_4'$ relative to the bodies $m_1'$ and $m_2'$ based on the value of $m''$, where the four bodies can form a deltoid-type central configuration containing the Earth--Moon axis (see Figure~\ref{fig:case_d}). The role of the  In Figures~\ref{fig:m_convex}--\ref{fig:m_s_concave} the curves $E_i$ and $S_i$, coloured according to the value of $m''$, denote the locations of the bodies $m_3'$ and $m_4'$, respectively.

In Figure~\ref{fig:m_convex}, the curves $E_3$ and $S_3$ represent the locations of the bodies $m_3'$ and $m_4'$, respectively, that together with the Earth ($m_1'$) and Moon ($m_2'$) form a convex central configuration. The curves $E_i$ and $S_i$ for $i=\{1,2\}$ represent the locations of the bodies $m_3'$ and $m_4'$, respectively, leading to concave configurations when the inner body of the configuration is the Moon ($m_2'$) (see Figure~\ref{fig:m_m_concave}), and for $i=\{4,5\}$ when the inner body is the body $m_4'$ (see Figure~\ref{fig:m_s_concave}). The masses $m_1$ and $m_2$ defined by Equation~(\ref{eq:m_mass}) are the same at the endpoints of the curves $E_i$ and $S_i$. When $m''=0$, then $m_1=0$ and $m_2=0.0061$, and when $m''=1$, then $m_1=0.3320$ and $m_2=0.0041$. The calculated angles $\alpha$ and $\beta$ of the endpoints of curves $E_i$ and $S_i$ are summarized in Appendix \ref{app:caseC}.

\newtext{To conclude, for any fixed value of $m''$ there exists one convex deltoid-type configuration, whereas the number of concave configurations depends on the value of $m''$ and may vary within the range discussed above.}

\subsection{\newtext{Physical Interpretation and Scope}}

\newtext{We conclude this section with a brief remark on its scope and interpretation. The purpose of Section~\ref{sec:cc_em} is not to provide a complete dynamical study of the Earth--Moon four-body problem, nor to develop a detailed mission-design application. Rather, its role is to demonstrate how the framework introduced in Section~\ref{sec:overview} can be applied to a physically relevant system with prescribed mass ratios.}

\newtext{From this perspective, the configurations presented here may be viewed as generalized equilibrium structures of a symmetric Earth--Moon four-body setting, extending the classical libration-point concept to a four-body context. In this sense, the present study is related to earlier work on specific solutions of the four-body and restricted four-body problems, as well as to studies of the topology of $n$-body systems~\citep{moeckel1990on}. The contour-map representation used here is based on the equations developed in~\citet{erdi2016central} and~\citet{czirjak2019study}, and makes it possible to display how the number and geometry of the admissible configuration families depend on the mass of the additional body or bodies.}

\newtext{The conceptual motivation for considering such configurations is also connected with the role of libration points in the three-body problem and in astrodynamical \mbox{applications~\citep{szebehely1967theory,xu2016survey}.} At the same time, questions of stability, local dynamics, and orbit design near the configurations identified here lie beyond the scope of the present paper. We, therefore, regard the Earth--Moon application primarily as a physically motivated case study that illustrates the usefulness of the semi-analytical classification developed in Section~\ref{sec:overview} and may serve as a basis for future dynamical investigations.}

\section{Conclusions}
\label{sec:conclusion}

Central configurations represent fundamental equilibrium structures of the Newtonian $n$-body problem and play a key role in understanding the geometry and dynamics of gravitational systems. Despite their importance, the systematic characterization and enumeration of central configurations remain incomplete in the four-body case. In particular, determining the number and structure of symmetric configurations for arbitrary mass combinations has remained a challenging problem, due to the nonlinear coupling between mass parameters and geometric constraints.

In this work, we developed a semi-analytical framework for studying symmetric four-body Dziobek configurations (S4BDC). Building upon our previous results~\citep{erdi2016central,czirjak2019study}, we formulated a unified approach that enables the explicit determination of the number of central configurations directly from the mass parameters. \newtext{The key strength of the presented model lies in its ability to determine the number of admissible configurations directly from the mass parameters, without requiring prior knowledge of the geometric arrangement or a separate extremum analysis with place-specific masses.}

Using this framework, we demonstrated that symmetric four-body Dziobek configurations can be fully characterized in terms of their mass parameters. In particular, while isosceles trapezoidal and convex deltoid configurations are uniquely determined by the masses, concave deltoid configurations exhibit a richer structure, with the number of admissible configurations varying depending on the mass ratios. The presented semi-analytical tools allow these cases to be distinguished and classified explicitly, thereby providing a complete description of this class of symmetric central configurations.

As a physically relevant application, we applied the developed framework to the Earth--Moon system. Since central configurations correspond to equilibrium solutions in rotating reference frames, they generalize the concept of libration points to multi-body systems. By analyzing several mass arrangements involving Earth- and Moon-mass bodies, we determined the possible equilibrium configurations and identified continuous families of solutions whose geometry depends on the mass of an additional body. These results extend the classical equilibrium structures known from the restricted three-body problem to the four-body case and provide a more general description of equilibrium configurations in realistic gravitational environments.

Beyond the specific example considered here, the presented semi-analytical framework provides a general method for studying symmetric central configurations in four-body systems. Such configurations are closely related to the topology of the phase space and play an important role in determining the dynamical structure of gravitational systems. They may contribute to understanding the formation and stability of natural multi-body systems, including planetary systems, satellite systems, and exoplanetary architectures. Furthermore, since equilibrium configurations serve as organizing centers of nearby dynamical motion, the results may also support future investigations of spacecraft dynamics and mission design in multi-body environments.

\textls[-25]{The semi-analytical approach developed in this study opens several directions for future research. One natural extension is the investigation of the stability properties of the identified configurations and their associated invariant manifolds. Another important direction is the extension of the method to more general configurations, including asymmetric or spatial central configurations. In addition, applying the presented framework to other astrophysical systems, such as Sun--planet--satellite configurations or multi-planet systems, may provide further insight into the equilibrium structures and dynamical architecture of gravitational systems.}

In summary, the results presented here contribute to the semi-analytical understanding of symmetric central configurations in the four-body problem and provide a systematic framework for determining and classifying these configurations based solely on mass parameters. By bridging semi-analytical theory and physically relevant applications, this work provides a foundation for further studies of equilibrium structures and dynamical organization in multi-body gravitational systems.

\vspace{--6pt}




\authorcontributions{\textls[-15]{Conceptualization, Z.C., B.É. and E.F.-D.; methodology,  Z.C. and B.É.; software,  Z.C.; validation, Z.C., B.É. and E.F.-D.; formal analysis, Z.C. and B.É.; investigation, Z.C., B.É. and E.F.-D.; resources, Z.C., B.É. and E.F.-D.; data curation, Z.C.; writing---original draft preparation, Z.C., B.É. and E.F.-D.; writing--review and editing, Z.C., B.É. and E.F.-D.; visualization, Z.C.; supervision, B.É. and E.F.-D. All authors have read and agreed to the published version of the manuscript}.}

\funding{\textls[-35]{This project has received funding from the HUN-REN Hungarian Research Network. E.F-D. also received funding from the NKFIH excellence grant TKP2021-NKTA-64 and NKFIH Grant OTKA K-147131.}}

\dataavailability{The original contributions presented in this study are included in the article. Further inquiries can be directed to the corresponding authors.}

\conflictsofinterest{The authors declare no conflicts of interest.} 



\appendixtitles{yes} 
\appendixstart
\appendix
\section{The \texorpdfstring{\boldmath{$a_0$}}{a0}, \texorpdfstring{$a_1$}{a1}, \texorpdfstring{$b_0$}{b0}, \texorpdfstring{$b_1$}{b1} Coefficients}
\label{app:coeff}
Convex case:\vspace{-6pt}
\begin{equation*}
\begin{split}
a_0 &= \tan\alpha\left(\cos^3\alpha-\frac{1}{8}\right), \\
b_0 &= \tan\beta\left(\cos^3\beta-\frac{1}{8}\right), \\
a_1 &= \frac{1}{(\tan\alpha+\tan\beta)^2}+ \tan\beta\left(\frac{1}{8}-\cos^3\alpha-\cos^3\beta\right) - \frac{\tan\alpha}{8},\\
b_1 &= \frac{1}{(\tan\alpha+\tan\beta)^2}+ \tan\alpha\left(\frac{1}{8}-\cos^3\alpha-\cos^3\beta\right) - \frac{\tan\beta}{8},
\end{split}
\end{equation*}
\par
\noindent
\vspace{-6pt}Concave case:
\begin{equation*}
\begin{split}
a_0 &= \tan\alpha\left(\cos^3\alpha-\frac{1}{8}\right), \\
b_0 &= -\tan\beta\left(\cos^3\beta-\frac{1}{8}\right), \\
a_1 &= \frac{1}{(\tan\alpha-\tan\beta)^2}+ \tan\beta\left(\frac{1}{8}-\cos^3\alpha-\cos^3\beta\right) - \frac{\tan\alpha}{8},\\
b_1 &= \frac{1}{(\tan\alpha-\tan\beta)^2}+ \tan\alpha\left(\frac{1}{8}-\cos^3\alpha-\cos^3\beta\right) + \frac{\tan\beta}{8},
\end{split}
\end{equation*}

\textls[-25]{Note that there are misprints in the coefficients $a_1$ and $b_1$ in the appendix of the paper~\citep{czirjak2019study}.}

\section{The Solution Counting Function for the Symmetric Deltoid Shaped Concave Case}
\label{app:counting}

\begin{figure}[H]

\hspace{-20pt}\includegraphics[width=0.7\linewidth]{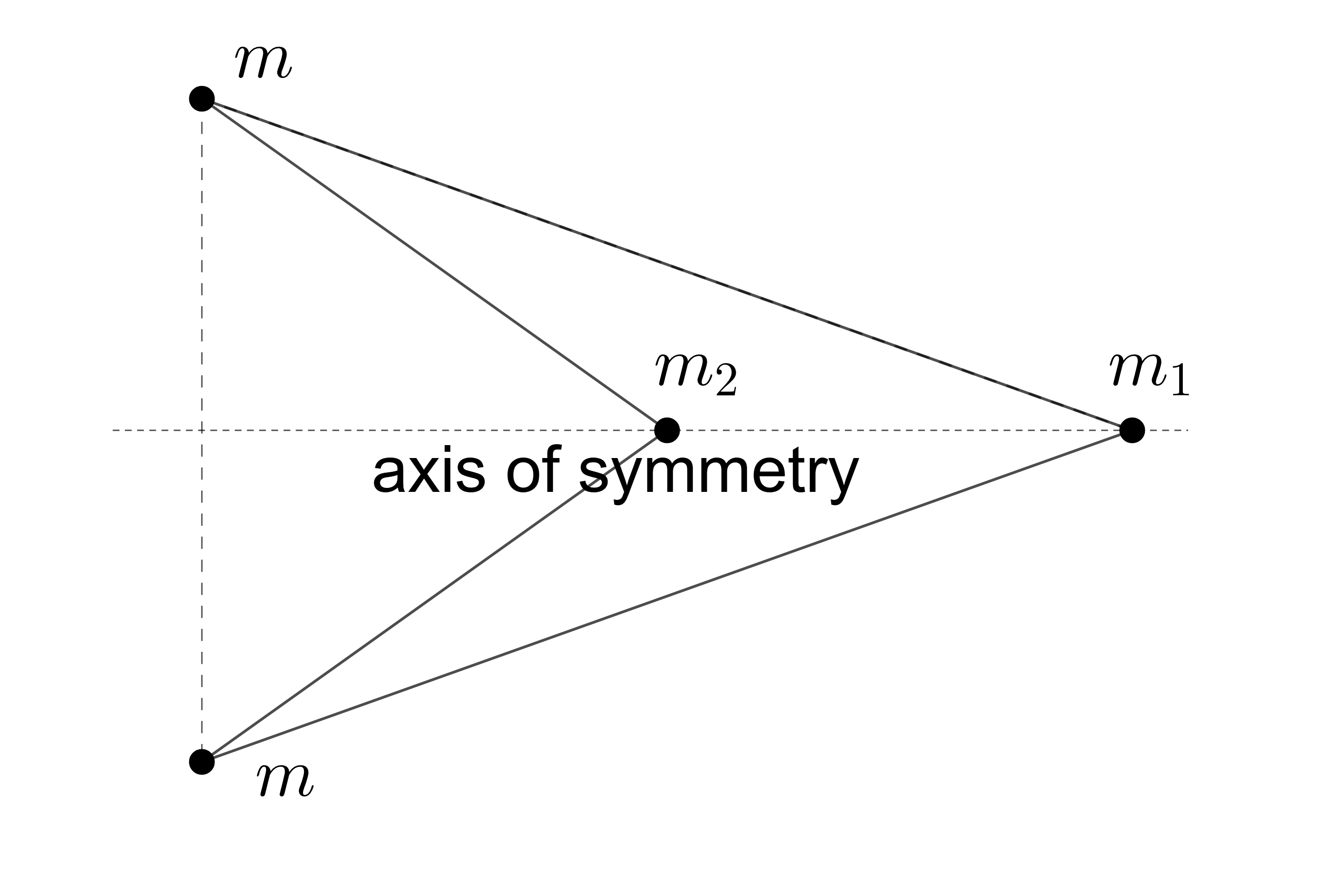}
\captionof{figure}{Concave deltoid type configuration where bodies with masses $m_1$ and $m_2$ are on the axis of symmetry. Bodies separated by the axis of symmetry have the same mass $m$.}
\label{fig:concave_deltoid}
\end{figure}

\begin{equation*}
N(m_1,m_2) = 
\begin{cases}
0 & m_1 > M_{m_2} \text{ and } m_2 > M_{m_1},\\
1 & \begin{array}{l}
	 m_1 > M_{m_2} \text{ and } m_2 = M_{m_1} \text{ or}\\
	 m_1 = M_{m_2} \text{ and } m_2 > M_{m_1} \text{ or}\\
	 m_1 = M_{m_2} \text{ and } m_2 = M_{m_1},
	\end{array}\\
2 & \begin{array}{l}
	 m_1 < M_{m_2} \text{ and } m_2 > M_{m_1} \text{ or}\\
	 m_1 > M_{m_2} \text{ and } m_2 < M_{m_1} \text{ or}\\
	 m_1 < M_{m_2} \text{ and } m_2 < M_{m_1} \text{ and } m_1=m_2,
	\end{array}\\
3 & \begin{array}{l}
	 m_1 < M_{m_2} \text{ and } m_2 = M_{m_1} \text{ or}\\
	 m_1 = M_{m_2} \text{ and } m_2 < M_{m_1},
	\end{array}\\
4 & m_1 < M_{m_2} \text{ and } m_2 < M_{m_1} \text{ and } m_1 \neq m_2,
\end{cases}
\end{equation*}
where $m_1+m_2 + 2m=1$, $M_{m_1} = M_1(\mu_2=m_1)$ and $M_{m_2} = M_1(\mu_2=m_2)$. The $M_1(\mu_2)$ function is approximated with the expression (\ref{eq:approx_M1}).

\section{Curve Parameter Table for Case \ref{em:b}}
\vspace{-6pt}
\label{app:caseB}
\begin{table}[H]
\scriptsize
 
 \caption{{The} 
 configurations parameters calculated in Case \ref{em:b} (see Figure~\ref{fig:case_b}) at the endpoints of the curves $S_i$ and $S_i'$ in Figure~\ref{fig:em_map_big}.}
 \label{tab:em_param}
\begin{tabularx}{\textwidth}{|CCCCCCC|}
 \hline
 \multicolumn{2}{|c}{\textbf{Positions}} & \multicolumn{5}{c|}{\textbf{Parameters}} \\
 \hline
 \textbf{Curves} & \textbf{Points} & \boldmath{$m''$} & \boldmath{$\alpha$} & \boldmath{$\beta$} & \boldmath{$m_1$} & \boldmath{$m_2$} \\
 \hline
  \multirow{2}{*}{$S_{1a}$ and $S_{1a}'$} & starting points $P_1$ and $P_1'$ & $\nu_1=0.0111$ & $65.19^\circ$ & $40.42^\circ$ & $0.9667$ & $0.0119$ \\ 
 & ending points & $1$ & $48.79^\circ$ & $0.18^\circ$ & $0.3320$ & $0.0041$ \\
 \hline
 \multirow{2}{*}{$S_{1b}$ and $S_{1b}'$} & starting points $P_1$ and $P_1'$ & $\nu_1=0.0111$ & $65.19^\circ$ & $40.42^\circ$ & $0.9667$ & $0.0119$ \\ 
 & ending points & $1$ & $71.12^\circ$ & $59.79^\circ$ & $0.3320$ & $0.0041$ \\
 \hline
 \multirow{2}{*}{ $S_{2a}$ and $S_{2a}'$} & starting points $P_2$ and $P_2'$ & $\nu_2=0.7114$ & $59.66^\circ$ & $16.24^\circ$ & $0.4107$ & $0.0051$ \\ 
 & ending points & $1$ & $59.62^\circ$ & $6.14^\circ$ & $0.3320$ & $0.0041$ \\
 \hline
 \multirow{2}{*}{$S_{2b}$ and $S_{2b}'$} & starting points $P_2$ and $P_2'$ & $\nu_2=0.7114$ & $59.66^\circ$ & $16.24^\circ$ & $0.4107$ & $0.0051$ \\ 
 & ending points & $1$ & $59.99^\circ$ & $29.99^\circ$ & $0.3320$ & $0.0041$ \\
 \hline
 \multirow{2}{*}{$S_3$ and $S_3'$} & starting points & 0 & $30^\circ$ & $30^\circ$ & $0.9878$ & $0.0122$ \\ 
 & ending points & $1$ & $59.82^\circ$ & $52.19^\circ$ & $0.3320$ & $0.0041$ \\
 \hline
 \end{tabularx}
\end{table}

\section{Curve Parameter Table for Case \ref{em:c}}
\vspace{-6pt}
\label{app:caseC}

\begin{table}[H]
 \footnotesize
 \caption{{The}  configurations parameters calculated in Case \ref{em:c} (see Figure~\ref{fig:case_c}) at the endpoints of the curves $S_i$ and $M_i$ in Figures~\ref{fig:e_convex} and \ref{fig:e_concave}.}
 \label{tab:e_param}
\begin{tabularx}{\textwidth}{|CCcCCcc|}
 \hline
 \multicolumn{2}{|c}{\textbf{Positions}} & \multicolumn{5}{c|}{\textbf{Parameters}} \\
 \hline
 \textbf{Curves} & \textbf{Points} & \boldmath{$m''$} & \boldmath{$\alpha$}\textbf{[deg]} & \boldmath{$\beta$}\textbf{[deg]} & \boldmath{$m_1$} & \boldmath{$m_2$} \\
 \hline
  \multirow{2}{*}{$S_1$ and $M_1$} & starting points & $0$ & $59.99^\circ$ & $15.77^\circ$ & $0.9759$ & $0.012$ \\ 
 & ending points & $1$ & $30.11^\circ$ & $30.11^\circ$ & $0.4959$ & $0.006$ \\
 \hline
 \multirow{2}{*}{$S_2$ and $M_2$} & starting points & $0$ & $45.05^\circ$ & $0.01^\circ$ & $0.9759$ & $0.012$ \\ 
 & ending points $P_{m_3'}$ and $P_{m_4'}$ & $\nu=0.0137$ & $65.18^\circ$ & $40.41^\circ$ & $0.0132$ & $0.9631$ \\
 \hline
 \multirow{2}{*}{$S_3$ and $M_3$} & starting points & 0 & $74.9^\circ$ & $59.99^\circ$ & $0.976$ & $0.012$ \\ 
 & ending points $P_{m_3'}$ and $P_{m_4'}$ & $\nu=0.0137$ & $65.18^\circ$ & $40.41^\circ$ & $0.0132$ & $0.9631$ \\
 \hline 
 \end{tabularx}
\end{table}

\section{Curve Parameter Table for Case \ref{em:d}}
\label{app:caseD}

\vspace{-6pt}\begin{table}[H]
 
 \caption{{The}  configurations angle parameters $\alpha$ and $\beta$ calculated in Case \ref{em:d} (see Figure~\ref{fig:case_d})  at the endpoints of the curves $E_i$ and $S_i$.}
 \label{tab:m_param}
\begin{tabularx}{\textwidth}{|CCCC|}
 \hline
 \multicolumn{2}{|c}{\textbf{Positions}} & \multicolumn{2}{c|}{\textbf{Parameters}} \\
 \hline
 \textbf{Curves} & \textbf{Points} & \boldmath{$\alpha$}\textbf{[deg]} & \boldmath{$\beta$}\textbf{[deg]}\\
 \hline
  \multirow{2}{*}{$S_1$ and $E_1$} & starting points & $63.63^\circ$ & $59.99^\circ$ \\ 
 & ending points &  $59.99^\circ$ & $29.99^\circ$ \\
 \hline
 \multirow{2}{*}{$S_2$ and $E_2$} & starting points & $59.57^\circ$ & $0.01^\circ$ \\ 
 & ending points & $59.63^\circ$ & $6.14^\circ$ \\
 \hline
 \multirow{2}{*}{$S_3$ and $E_3$} & starting points & $59.99^\circ$ & $59.89^\circ$ \\ 
 & ending points & $59.82^\circ$ & $52.82^\circ$ \\
 \hline
 \multirow{2}{*}{$S_4$ and $E_4$} & starting points & $60^\circ$ & $0.05^\circ$ \\ 
 & ending points & $48.79^\circ$ & $0.18^\circ$ \\
 \hline
 \multirow{2}{*}{$S_5$ and $E_5$} & starting points & $60^\circ$ & $55.55^\circ$ \\ 
 & ending points & $71.12^\circ$ & $59.79^\circ$ \\
 \hline
 \end{tabularx}
\end{table}

\begin{adjustwidth}{-\extralength}{0cm}

\reftitle{References}



\PublishersNote{}
\end{adjustwidth}
\end{document}